\documentclass[preprint]{aastex631}

\usepackage{xspace}

\usepackage{amsmath}

\newcommand\RN{2015 RN$_{35}$\,}

\newcommand\JDzero{2459932.5~JD}
\newcommand\Nmc{1000\xspace}
\newcommand\Pone{1149.7}
\newcommand\Poneerr{$1149.7\pm0.3$\xspace}
\newcommand\Ptwo{896.01}
\newcommand\Ptwoerr{$896.01\pm0.01$\xspace}

\newcommand\gcenter{0.481~$\mu$m\xspace}
\newcommand\gr{$0.714\pm0.008$\xspace}
\newcommand\ri{$0.245\pm0.009$\xspace}
\newcommand\rz{$0.255\pm0.020$\xspace}

\newcommand\HJPL{23.24\xspace}
\newcommand\HV{23.9$\pm$0.2\xspace}
\newcommand\Gone{$-0.10\pm0.08$\xspace}
\newcommand\Gtwo{$0.8\pm0.1$\xspace}
\newcommand\diam{$41\pm8$\xspace}
\newcommand{\degdeg}{^{\circ}}

\received{2023 August 25}
\accepted{2023 October 7}

\submitjournal{AJ}

\shortauthors{Beniyama et al.}
\graphicspath{{./}{../plot/fig/}{../plot/tab}}

\turnoffediting

\begin{document}

\title{
Multicolor Photometry of Tiny Near-Earth Asteroid 2015 RN$_{35}$ Across a Wide Range of Phase Angles: Possible Mission Accessible A-type Asteroid
\footnote{Created on January 15th, 2023}
}
\shorttitle{2015 RN$_{35}$}


\correspondingauthor{Jin Beniyama}
\email{beniyama@ioa.s.u-tokyo.ac.jp}
\author[0000-0003-4863-5577]{Jin Beniyama}
\affiliation{%
Institute of Astronomy, Graduate School of Science,\\
The University of Tokyo, 2-21-1 Osawa, Mitaka, Tokyo 181-0015, Japan}
\affiliation{%
Department of Astronomy, Graduate School of Science,\\
The University of Tokyo, 7-3-1 Hongo, Bunkyo-ku, Tokyo 113-0033, Japan}
\author[0000-0003-4814-0101]{Ryou Ohsawa}
\affiliation{%
National Astronomical Observatory of Japan, 2-21-1 Osawa, Mitaka, Tokyo 181-8588, Japan}
\author[0000-0001-8228-8789]{Chrysa Avdellidou}
\affiliation{%
Université Côte d'Azur, 
Observatoire de la Côte d'Azur, CNRS, Laboratoire Lagrange, Bd de l'Observatoire, 
CS 34229, 06304 Nice Cedex 4, France}
\author[0000-0002-8792-2205]{Shigeyuki Sako}
\affiliation{%
Institute of Astronomy, Graduate School of Science,\\
The University of Tokyo, 2-21-1 Osawa, Mitaka, Tokyo 181-0015, Japan}
\affiliation{%
UTokyo Organization for Planetary Space Science, 
The University of Tokyo, 7-3-1 Hongo, Bunkyo-ku, 
Tokyo 113-0033, Japan}
\affiliation{%
Collaborative Research Organization for
Space Science and Technology, 
The University of Tokyo, 7-3-1 Hongo, 
Bunkyo-ku, Tokyo 113-0033, Japan}
\author{Satoshi Takita}
\affiliation{%
Institute of Astronomy, Graduate School of Science,\\
The University of Tokyo, 2-21-1 Osawa, Mitaka, Tokyo 181-0015, Japan}
\author[0000-0002-7332-2479]{Masateru Ishiguro}
\affiliation{%
Department of Physics and Astronomy, 
Seoul National University, 1 Gwanak-ro, Gwanak-gu, 
Seoul 08826, Republic of Korea}
\affiliation{%
SNU Astronomy Research Center, 
Seoul National University, 
1 Gwanak-ro, Gwanak-gu, Seoul 08826, Republic of Korea}
\author[0000-0003-1726-6158]{Tomohiko Sekiguchi}
\affiliation{%
Asahikawa Campus, Hokkaido University of Education, 
Hokumon 9, Asahikawa, Hokkaido 070-8621, Japan}
\author[0000-0003-2273-0103]{Fumihiko Usui}
\affiliation{%
Institute of Space and Astronautical Science,
Japan Aerospace Exploration Agency, 3-1-1 Yoshinodai,
Chuo-ku, Sagamihara, Kanagawa 252-5210, Japan}
\author[0000-0001-5456-4977]{Shinichi W. Kinoshita}
\affiliation{%
Department of Astronomy, Graduate School of Science,\\
The University of Tokyo, 7-3-1 Hongo, Bunkyo-ku, Tokyo 113-0033, Japan}
\affiliation{%
National Astronomical Observatory of Japan, 2-21-1 Osawa, Mitaka, Tokyo 181-8588, Japan}
\author[0000-0003-4814-0101]{Kianhong Lee}
\affiliation{%
Institute of Astronomy, Graduate School of Science,\\
The University of Tokyo, 2-21-1 Osawa, Mitaka, Tokyo 181-0015, Japan}
\affiliation{%
National Astronomical Observatory of Japan, 2-21-1 Osawa, Mitaka, Tokyo 181-8588, Japan}
\affiliation{%
Astronomical Institute, Tohoku University, Sendai 980-8578, Japan}
\author{Asami Takumi}
\affiliation{%
Open University of Japan, 2-11 Wakaba, Mihama-ku, Chiba 261-8586, Japan}
\affiliation{%
National Astronomical Observatory of Japan, 
2-21-1 Osawa, Mitaka, Tokyo 181-8588, Japan}
\author[0000-0002-0535-652X]{Marin Ferrais}
\affiliation{%
Arecibo Observatory, University of Central Florida, HC-3 Box 53995, Arecibo, PR 00612, USA}
\author[0000-0001-8923-488X]{Emmanu{\"e}l Jehin}
\affiliation{%
Space sciences, Technologies and Astrophysics Research (STAR) Institute, 
Université de Liège, Allée du 6 Août 17, 4000 Liège, Belgium}


\begin{abstract}
Studying small near-Earth asteroids is 
important to understand their dynamical histories and origins as well as
to mitigate the damage of the asteroid impact to the Earth.
We report the results of multicolor photometry of the tiny near-Earth asteroid 2015 RN$_{35}$ 
using the 3.8~m Seimei telescope in Japan and the TRAPPIST-South telescope in Chile over 17 nights in 2022 December and 2023 January.
We observed 2015 RN$_{35}$ across a wide range of phase angles from 
2$^{\circ}$ to 30$^{\circ}$ in the $g$, $r$, $i$, and $z$ bands in the Pan-STARRS system.
These lightcurves show that 2015 RN$_{35}$ is in a non-principal axis spin state
with two characteristic periods of \Poneerr~s and \Ptwoerr~s.
We found that a slope of a visible spectrum of 2015 RN$_{35}$ 
is as red as asteroid (269) Justitia, one of the very red objects in the main belt,
which indicates that 2015 RN$_{35}$ can be classified as an A- or Z-type asteroid.
In conjunction with the shallow slope of the phase curve,
we suppose that 2015 RN$_{35}$ is a high-albedo A-type asteroid.
We demonstrated that surface properties of tiny asteroids could be well constrained by 
intensive observations across a wide range of phase angles.
2015 RN$_{35}$ is a possible mission accessible A-type near-Earth asteroid with a small $\Delta v$ of 11.801~km\,s$^{-1}$ 
in the launch window between 2030 and 2035.
\end{abstract}


\section{Introduction}\label{sec:intro}
It is now well established that the first stage of the planetary formation process is the accretion of the so-called planetesimals from the solids in our protoplanetary disk. 
Theoretical and observational studies have shown that the planetesimals were formed at large sizes, diameters ($D$) larger than 50 to 100~km \citep{Delbo2017}. 
The small bodies of our solar system are remnants of that era. 
However, not all the current asteroids are survivors from primordial times. 
Collisions between this original planetesimal population produced clusters of fragments of smaller sizes, the so-called asteroid families. 
A nongravitational effect—the Yarkovsky effect—slowly changes the orbital semimajor axis $a$ of asteroids at a rate d$a$/dt proportional to 1/$D$ \citep{Vokrouhlicky1998}. 
Asteroids in prograde rotation have d$a$/d$t > 0$ and migrate towards larger heliocentric distances, 
whereas those in retrograde rotation with d$a$/d$t < 0$ migrate towards the Sun. 
Another effect that is also caused by the solar radiation, the Yarkovsky-O'Keefe-Radzievskii-Paddack (YORP) effect \citep{Rubincam2000}, 
can change the spin state of asteroids affecting the rate of the drift due to Yarkovsky. 
Both the Yarkovsky and YORP effects depend on the surface properties of the asteroids and their internal structure.
The migration of small main belt asteroids can lead the smaller ones to reach the dynamical routes (resonances with planets) 
that can bring them to the near-Earth space, hence sampling several regions (as well as asteroid families) of the main belt.
Studying near-Earth asteroids (NEAs) is therefore crucial to understand the material transportation from the main belt to the near-Earth space
as well as to mitigate the hazard of an asteroid impact to the Earth.

Tiny asteroids having diameters less than 100~m could be characterized during 
their close approaches to the Earth using ground-based and space-borne telescopes.
Comprehensive studies of large asteroids have been conducted, whereas 
only few studies focus on tiny asteroids due to observational difficulties caused by
limited visibilities and large apparent motions of asteroids during their close approaches.

Using the Infrared Array Camera (IRAC) on the Spitzer Space Telescope,
\cite{Mommert2014a, Mommert2014b} conducted infrared observations of 
tiny NEAs 2009 BD ($D\leq5$ m) and 2011 MD ($D\sim10$ m).
According to these observations, asteroid 2009 BD has inconclusive surface 
nature which could be either covered by fine regolith or composed of a collection of bare rocks,
while the bulk density of 2011 MD is estimated 
to be $1.1^{+0.7}_{-0.5}\times10^3\,\mathrm{kg\,m^{-3}}$, indicating a rubble-pile origin.
Recently, \cite{Fenucci2021, Fenucci2023b} found that the tiny superfast rotators
(499998) 2011 PT ($D\sim35$ m and rotation period $P\sim10$ minutes) and 2016 GE$_1$ ($D\sim12$ m and $P\sim34$ s) have
small thermal conductivities of $K\leq0.1\,\mathrm{W\,m^{-1} K^{-1}}$ and $K\leq100\,\mathrm{W\,m^{-1}K^{-1}}$, respectively.
Such small conductivities imply that 
these two tiny asteroids are covered with fine regolith or highly porous rocks \citep{Avdellidou2020, Cambioni2021}.
On the other hand,
a bunch of fast rotators are discovered in video observations using a CMOS camera \citep{Beniyama2022}. 
Some of them need to have strength similar to the typical tensile strength of meteorites
to keep their fast rotations. It is unclear that such fast rotators could have fine regolith on
its surface.
Thus, it is still in debate whether tiny asteroids are monolithic or rubble-pile, and with or without fine regolith on their surface.

\cite{Reddy2016} conducted radar, lightcurve, and spectroscopic observations
of the tiny E-type NEA 2015 TC$_{25}$ ($D\sim2$ m).
They concluded that 2015 TC$_{25}$ is a fragment possibly ejected from the E-type main belt asteroid (44) Nysa.
One of the interesting properties of 2015 TC$_{25}$ is its bluer spectrum compared 
to a typical E-type.
They explained the bluer slope of 2015 TC$_{25} $ in the visible wavelength with a lack of fine regolith on the surface due to a combination of 
weak gravity and fast rotation. 
Recently, \cite{Licandro2023} found that a visible spectrum of 
the tiny fast-rotating asteroid 2022 AB ($D\sim65$ m and $P\sim3$ minutes) 
shows an upturn over the 0.4 to 0.6~$\mu$m, which does not fit with any known asteroid spectrum
\footnote{
    We note that the spectrum is similar to that of the Martian Trojan (121514) 1998 UJ$_7$ \citep{Borisov2018b}.
}.

The phase angle dependence of an asteroid brightness, the so-called phase curve, 
informs about the surface properties of the asteroid \citep[see, e.g.;][]{Bowell1989, Belskaya2000}.
High-albedo asteroids have shallower slope in their phase curves 
since the contribution of the shadow-hiding decreases as albedo increases \citep{Belskaya2000}, 
whereas low-albedo asteroids have steeper slopes.
Apart from the albedo, other properties are related to the phase curve
such as the surface grain size and roughness.
An important consideration to study phase curves of asteroids is the rotation correction \citep[see, e.g.;][]{Harris1989}. 
Homogeneous sets of the typical brightness such as maximum and mean of the lightcurves at certain phase angles are necessary 
to accurately derive the related quantities,
otherwise, the brightness variation caused by the rotation leads to misunderstanding of observational results.
Thus, the tiny asteroids, which are often fast-rotating \citep{Thirouin2016, Thirouin2018, Beniyama2022} 
and do not require a long time to obtain a mean brightness across a rotation phase, are 
appropriate targets to investigate phase curves.

Well-sampled phase curves of small asteroids are
less commonly obtained since their observation opportunities are limited.
\cite{Reddy2015} characterized the small NEA 2004 BL$_{86}$ ($D\sim$300~m) at a wide phase angle range from 1.5$\degdeg$ to 49.6$\degdeg$.
The visible geometric albedo of about 0.4 derived from the slope of the phase curve
is consistent with the near-infrared spectrum of 2004 BL$_{86}$, 
which implies that 2004 BL$_{86}$ is a typical high-albedo V-type asteroid.
Recently, several dozens of phase curves of NEAs are studied in the framework of the IMPACTON project \citep{Rondon2019, Rondon2022, Ieva2022, Arcoverde2023}. 
They made a database with phase curves of 30 NEAs using three 1~m-class telescopes in Brazil and Italy \citep{Arcoverde2023}.
Their sample include only one tiny NEA, 2017 DC$_{38}$, 
with the absolute magnitude $H$ of 24.22,
that was successfully observed at a very small phase angle of 1.1$\degdeg$.
However, its rotation period was not obtained and thus no rotation correction has been performed in the analysis of the phase curve of 2017 DC$_{38}$.

In this paper, we present the results of multicolor photometry of the tiny NEA
\RN over 17 nights in Japan and Chile.
The target asteroid \RN was discovered by the Pan-STARRS 1 survey \citep{Chambers2016} on 2015 September 9.
\RN is an Apollo-class NEA, and its
absolute magnitude in the V band is \HJPL in NASA JPL Small-Body DataBase (SBDB)
\footnote{\url{https://ssd.jpl.nasa.gov/tools/sbdb_lookup.html}, last accessed 2023 August 10.}.
The trajectories of \RN were well studied and possible collisions with Earth were discussed \citep{Petrov2018}.
\RN had a close approach in 2022 December.
\RN was observable at phase angles from 30$\degdeg$ to 0.6 $\degdeg$ from 2022 December to 2023 January,
which is a quite rare opportunity to obtain a well-sampled phase curve of a tiny asteroid.
The paper is organized as follows.
In section \ref{sec:methods}, we summarize our observations and data reduction.
The physical properties of \RN are summarized in Section \ref{sec:result}.
The surface properties of the tiny asteroid \RN
and possible exploration by spacecraft mission are discussed in Section \ref{sec:discussion}.

\section{Observations and data reduction} \label{sec:methods}
We conducted photometric observations at two observatories in Japan and Chile.
The observing conditions are summarized in Table \ref{tab:obs}.
The predicted V band magnitudes, phase angles, distances between \RN and observer, and distances between \RN and the Sun in Table \ref{tab:obs} were obtained from 
{NASA JPL HORIZONS\footnote{\url{https://ssd.jpl.nasa.gov/horizons}}
using the \texttt{Python} package \texttt{astroquery} \citep{Ginsburg2019}.

\subsection{Seimei telescope} \label{subsec:seimei}
We obtained 12 lightcurves of \RN using 
TriColor CMOS Camera and Spectrograph (TriCCS) on
the 3.8~m Seimei telescope \citep{Kurita2020}
from 2022 December 23 to 2023 January 21.
We simultaneously took three-band images in the Pan-STARRS ($g$, $r$, $i$) and ($g$, $r$, $z$) filter \citep{Chambers2016}.
The field of view is $12.6\arcmin\times7.5\arcmin$ with a pixel scale of 0.350~arcsec/pixel.

Nonsidereal tracking was performed during the observations of \RN.
The exposure times were 5 or 60~s according to the brightness of 2015 RN$_{35}$.
The signal-to-noise ratios of \RN in the data taken in 2023 January are too low to detect \RN in a single exposure.
We took multiple images with short exposures rather than a single image with long exposures in our observations 
in order to avoid having elongated photometric reference stars and also to eliminate the cosmic rays.

We performed standard image reduction including bias subtraction, dark subtraction, and flat-fielding.
The astrometry of reference sources from the Gaia Data Release 2 was performed using the \texttt{astrometry.net} software \citep{Lang2010}.
For the data taken in 2023 January, we performed stacking of images before photometry to avoid the elongations of the images of \RN as shown in the upper panels of Figure \ref{fig:cutout}
(hereinafter referred to as the nonsidereally stacked image).
We stacked 20 successive images with exposure times of 60~s.
Since a typical readout time of the CMOS sensors on TriCCS is 0.4 milliseconds,
the total integration time is about 1200~s, which corresponds to
one of the characteristic periods of \RN (see Section \ref{subsec:res_lc}).
We also stacked images using the World Coordinate System (WCS) of images corrected with the surrounding sources
to suppress the elongations of the images of reference stars as shown in the lower panels of Figure \ref{fig:cutout}
(hereinafter referred to as the sidereally stacked image).

We derived colors and magnitudes of \RN following the same procedure described in \cite{Beniyama2023a, Beniyama2023b}.
Cosmic rays were removed with the \texttt{Python} package \texttt{astroscrappy} \citep{McCully2018}
using the Pieter van Dokkum's \texttt{L.A.Cosmic} algorithm \citep{vanDokkum2001}.
The circular aperture photometry was performed for 2015 RN$_{35}$ 
and the reference stars using the SExtractor-based \texttt{Python} package \texttt{sep} \citep{Bertin1996, Barbary2017}.
The aperture radii were set to twice as large as the full width at half maximums (FWHMs) of the point spread functions (PSFs) of 
the reference stars in the sidereally stacked images.
The photometric results of 2015 RN$_{35}$ and reference stars were obtained from 
the nonsidereal and sidereal stacked images, respectively.

\subsection{TRAPPIST-South telescope} \label{subsec:trappist}
We obtained five lightcurves of \RN using the robotic telescope
TRAPPIST-South \citep[the Minor Planet Center code I40;][]{Jehin2011}
of the University of Liège between 2022 December 19 and 26.
TRAPPIST-South is a 0.6-m Ritchey-Chrétien telescope operating at f/8 and equipped with a CCD camera FLI ProLine 3041-BB. 
The field of view is $22\arcmin\times22\arcmin$ with an un-binned pixel scale of 0.64~arcsec/pixel.

We obtained images in the sidereal tracking mode with the wide $Exo$-filter, 
whose wavelength coverage roughly corresponds to
the $r$ to $y$ bands in the Pan-STARRS system
\citep{Jehin2011}.
We set exposure time to 40~s on December 19, 20, 21, and 22 using the $2\times2$ binning mode, and to 120~s on December 26 while using no binning.

The raw images were processed using standard bias, dark and flat fields frames. 
The photometry was performed using the \texttt{PHOTOMETRYPIPELINE} \citep{Mommert2017} 
to derive the $r$ band magnitudes in the Pan-STARRS system. 
This pipeline allows zero-point calibration by matching field stars with online catalogs. 
Typically 100 stars with solar-like colors (i.e. stars with $g-r$ and $r-i$ colors closer than 0.2 magnitudes to that of the Sun) were used in each image for the magnitude calibration.
Aperture radii were set to 4~pixels for the binned observations and to 8~pixels for the un-binned mode.

\begin{deluxetable*}{lccccccccccc}
        \tablenum{1}
        \tablecaption{Summary of the observations\label{tab:obs}}
        \tablewidth{0pt}
        \tablehead{
            \colhead{Obs. Date} & Tel. & Filter & \colhead{$t_\mathrm{exp}$} & \colhead{$N_\mathrm{img}$} & V & $\alpha$ & $\Delta$ & $r_\mathrm{h}$ & Air Mass & Weather \\
            \colhead{(UTC)}     &      &         & \colhead{(s)}              &                            & (mag) & (deg) & (au) & (au) &   &
        }
        \decimals
        \startdata
        2022 Dec 19 05:05:40--08:40:18& TRAPPIST-S & $Exo$ & 40 & 233 & 15.3 & 31.1 &0.014 & 0.996 & 1.24--1.48 & Clear\\
2022 Dec 20 02:46:27--04:00:16& TRAPPIST-S & $Exo$ & 40 & 83 & 15.7 & 31.3 &0.017 & 0.998 & 1.71--2.64 & Clear\\
2022 Dec 21 07:13:33--08:34:49& TRAPPIST-S & $Exo$ & 40 & 91 & 16.1 & 31.1 &0.021 & 1.002 & 1.32--1.51 & Clear\\
2022 Dec 22 06:02:40--08:40:00& TRAPPIST-S & $Exo$ & 40 & 171 & 16.5 & 30.7 &0.024 & 1.004 & 1.33--1.55 & Clear\\
2022 Dec 23 16:32:56--17:14:41& Seimei & $g,r,i$ & 5 & 394 & 16.8 & 30.0 &0.029 & 1.008 & 1.07--1.09 & Clear\\
2022 Dec 23 19:03:49--19:45:32& Seimei & $g,r,z$ & 5 & 275 & 16.9 & 29.9 &0.029 & 1.009 & 1.20--1.33 & Clear\\
2022 Dec 26 05:04:08--08:39:14& TRAPPIST-S & $Exo$ & 120 & 100 & 17.4 & 28.1 &0.037 & 1.016 & 1.39--1.64 & Clear\\
2022 Dec 26 18:09:43--18:35:49& Seimei & $g,r,i$ & 30 & 53 & 17.5 & 27.7 &0.039 & 1.018 & 1.10--1.14 & Clear\\
2023 Jan 12 12:15:06--12:35:09& Seimei & $g,r,i$ & 1200 & 2 & 19.1 & 10.9 &0.100 & 1.082 & 1.52--1.66 & Cirrus\\
2023 Jan 16 14:54:23& Seimei & $g,r,i$ & 1200 & 1 & 19.3 & 6.7 &0.117 & 1.100 & 1.06--1.06 & Clear\\
2023 Jan 17 12:11:24& Seimei & $g,r,i$ & 1200 & 1 & 19.3 & 5.8 &0.121 & 1.104 & 1.53--1.53 & Cirrus\\
2023 Jan 18 14:07:50--14:30:15& Seimei & $g,r,i$ & 1200 & 2 & 19.4 & 4.7 &0.125 & 1.109 & 1.08--1.11 & Clear\\
2023 Jan 18 14:51:09& Seimei & $g,r,z$ & 1200 & 1 & 19.4 & 4.7 &0.126 & 1.109 & 1.06--1.06 & Clear\\
2023 Jan 20 14:04:41& Seimei & $g,r,i$ & 1200 & 1 & 19.4 & 2.8 &0.134 & 1.118 & 1.10--1.10 & Clear\\
2023 Jan 20 14:25:34& Seimei & $g,r,z$ & 1200 & 1 & 19.4 & 2.8 &0.134 & 1.118 & 1.07--1.07 & Clear\\
2023 Jan 21 14:03:12& Seimei & $g,r,i$ & 1200 & 1 & 19.4 & 1.9 &0.139 & 1.123 & 1.10--1.10 & Clear\\
2023 Jan 21 14:27:02--14:47:05& Seimei & $g,r,z$ & 1200 & 2 & 19.4 & 1.9 &0.139 & 1.123 & 1.05--1.07 & Clear\\
  \enddata
            \tablecomments{
            Observation time in UT in midtime of exposure (Obs. Date), telescope (Tel.), filters (Filters), 
            total exposure time per frame ($t_{\mathrm{exp}}$),
            the number of images ($N_\mathrm{img}$),
            and weather condition (Weather) are listed.
            Predicted V band apparent magnitude (V), 
            phase angle ($\alpha$),
            distance between 2015 RN$_{35}$ and observer ($\Delta$),
            and 
            distance between 
            2015 RN$_{35}$ and Sun ($r_\mathrm{h}$) at the observation starting time
            are referred to NASA Jet Propulsion Laboratory (JPL) HORIZONS
             as of 2023 August 19 (UTC).
            Elevations of 2015 RN$_{35}$ to calculate air mass range (Air Mass) are 
            also referred to NASA JPL HORIZONS.
            }
            \end{deluxetable*}

\begin{figure*}[ht]
\plotone{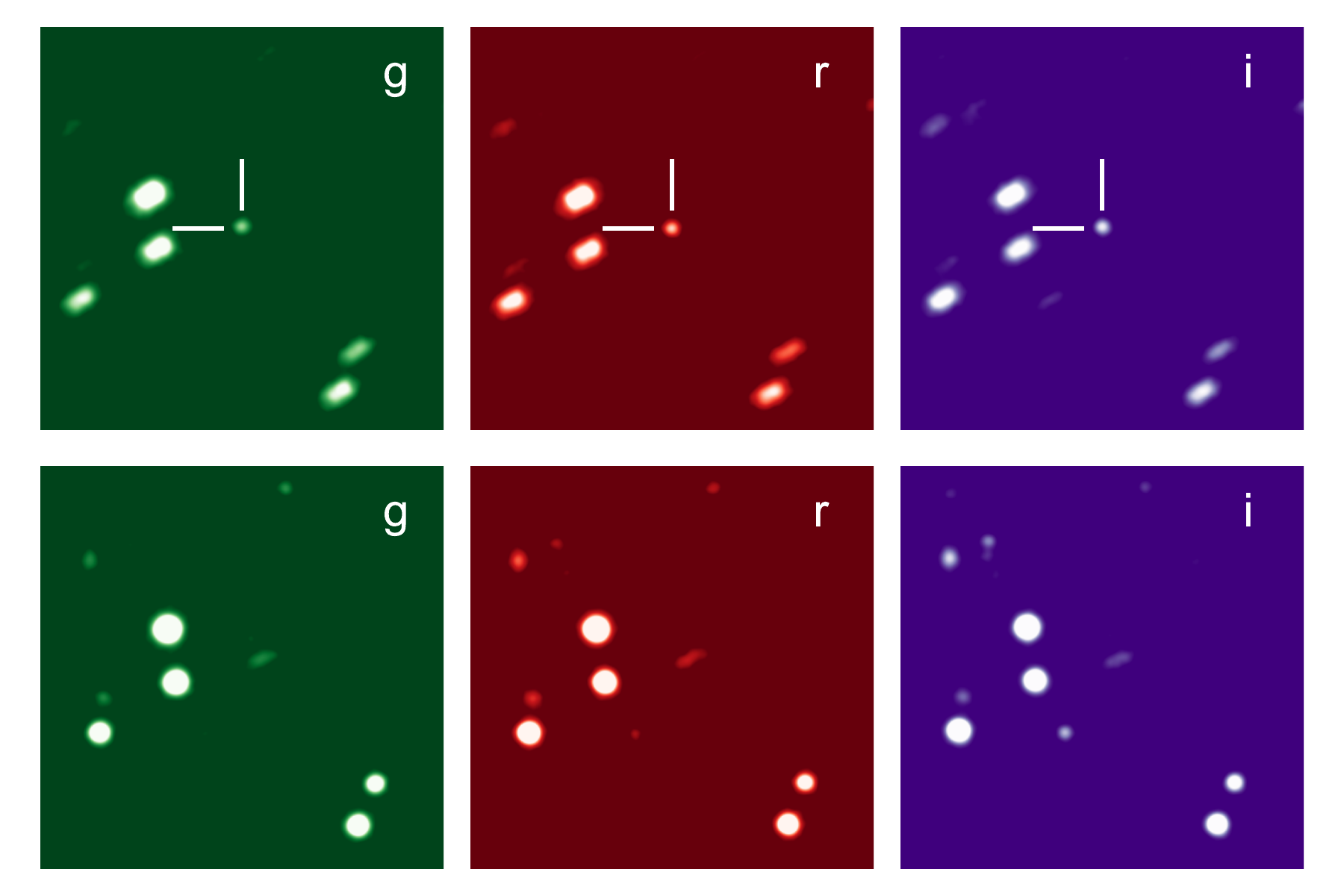}
\caption{
    Stacked images of 2015 RN$_{35}$ in $g$, $r$, and $i$ bands with a total integration time of 1200~s 
   in 2023 January 16. Nonsidereally stacked images (top) and sidereally stacked images (bottom) are shown.
   Horizontal and vertical bars indicate \RN.
   Field of view covers $1.75\arcmin\times1.75\arcmin$. North is to the top and East is to the left.
}
\label{fig:cutout}
\end{figure*}

\section{Results} \label{sec:result}
\subsection{Lightcurves and rotation period} \label{subsec:res_lc}
The light-travel time was corrected to obtain the time-series colors and magnitudes of \RN \citep{Harris1989}.
The eight lightcurves of 2015 RN$_{35}$ taken on 2022 December are shown in Figure \ref{fig:lc}.
Brightness variation of about 0.7~mag is seen in each lightcurve. 
The lightcurves show non-perfect periodic signals, implying that 2015 RN$_{35}$ is a non-principal axis rotator in a complex rotation state \citep[i.e., a tumbler;][]{Pravec2005}.

We performed the periodic analysis using the Lomb--Scargle technique \citep{Lomb1976, Scargle1982, VanderPlas2018}
with three long lightcurves obtained with the TRAPPIST-South telescope on 2022 December 19, 22, and 26. 
The Lomb--Scargle periodograms with a period range between 500 to 2000 s are shown in 
Figure \ref{fig:periodogram}, where four peak frequencies, $f_a$, $f_b$, $f_c$, and $f_d$, are indicated.
We focused on the two strongest frequencies, $f_b$ and $f_d$, based on the powers of periodograms.
The $f_d$ appears to be the first overtone of $f_b$: $2f_b \fallingdotseq f_d$.
We regarded that $f_b$ corresponds to a period of \RN, $P_1$, since folded lightcurves with $f_b^{-1}$ are typical double-peaked lightcurves.
The uncertainty of $P_1$ was estimated with the Monte Carlo technique following the previous work \citep{Beniyama2022}.
We obtained \Nmc lightcurves by randomly resampling the data 
assuming each observed data-point follows a normal distribution whose standard deviation is a photometric error.
We calculated the \Nmc periods corresponding to $P_1$ and derived it with the uncertainty each night as $1149.7\pm0.4$ s (Dec. 19), $1149.6\pm0.5$ s (Dec. 22), and $1149.9\pm0.5$ s (Dec. 26).
We adopted the error-weighted average of these three periods, \Poneerr, as $P_1$.

Figure \ref{fig:plc} shows the eight $r$ band lightcurves of 2015 RN$_{35}$ folded with the period of \Pone~s.
The folded lightcurves seem to be double-peaked but not perfectly overlapped each other in rotation phase, probably due to the non-principal axis rotation.
The model curve with the period of \Pone~s is also shown in Figure \ref{fig:plc}.
The root mean square of residual (RMS) is calculated as follows:
\begin{eqnarray}
    \mathrm{RMS} =  \sqrt{\sum_i^{n_\mathrm{obs}} \frac{\lbrack y_\mathrm{obs}(t_i) - y_\mathrm{model}(t_i)\rbrack^2}{n_\mathrm{obs}}}  
\end{eqnarray}
where $n_\mathrm{obs}$ is the number of observation data,
$t_i$ is the observation time of the $i$-th sample,
$y_\mathrm{obs}(t_i)$ is $i$-th observed brightness at $t_i$,
and $y_\mathrm{model}(t_i)$ is the model brightness at $t_i$.
The RMS is calculated as 0.104.

\cite{Franco2023b} derived the rotation period of 2015 RN$_{35}$ to be 
$0.3193\pm0.0001$ hr $\sim1149$ s using lightcurves obtained on 2022 December 18 and 19.
\cite{Kolenczuk2023} found the rotation period of
$19.1692\pm0.0069$ minutes $\sim1150$ s from the intensive observation campaign during 2022 December. 
These reported periods are close to $P_1$ and corresponding to $f_b$ in Figure \ref{fig:periodogram}.
On the other hand, \cite{Colazo2023} derived the rotation period of 2015 RN$_{35}$ to be 
$0.478\pm0.008$ hr $\sim1721$ s using lightcurves obtained on 2022 December 16,
which corresponds to $f_a$ in Figure \ref{fig:periodogram}.
Since the phased lightcurves in \cite{Colazo2023} appear as not double-peaked 
unlike others, the rotation period of $\sim1150$ is highly likely.

We continue the periodic analysis for the three lightcurves obtained with the TRAPPIST-South telescope on 2022 December 19, 22, and 26
following procedures in previous studies \citep{Pravec2005, Pravec2014, Lee2017, Lee2022}.
The purpose of successive analysis is to derive the other period $P_2$ characterizing the non-principal axis rotation of 2015 RN$_{35}$.
Searching for the $P_2$ is performed against the all five lightcurves obtained with the TRAPPIST-South telescope.
We fit the lightcurves with two-dimensional Fourier series keeping $P_1$ fixed:
\begin{align}
    &y_\mathrm{model}(t) = C_0 + \sum_{j=1}^m \left\lbrack C_{j0}\cos{\left(\frac{2\pi j}{P_1}t\right)} + S_{j0}\sin{\left(\frac{2\pi j}{P_1}t\right)}\right\rbrack \nonumber \\ \nonumber
                        &+ \sum_{k=1}^m \sum_{j=-m}^m \left\lbrack C_{jk}\cos{\left(\frac{2\pi j}{P_1}t+\frac{2\pi k}{P_2}t\right)} + S_{jk}\sin{\left(\frac{2\pi j}{P_1}t+\frac{2\pi k}{P_2}t\right)}\right\rbrack,
\end{align}
where $t$ is time, $m$ is the order of Fourier series, $C_0$, $C_{jk}$ and $S_{jk}$ are the Fourier coefficients.
We set $P_2$ and Fourier coefficients as free parameters and searched $P_2$ from 
100~s to 10000~s with a step of 1~s.
The RMS residual between observed and model lightcurves of 2015 RN$_{35}$ is calculated in each step.
The results of the search of $P_2$ are shown in Figure \ref{fig:rms}.
We plotted results of grid search of $P_2$ in cases with $m$ of 3 and 4.
Five periods with smaller RMSs in the two cases, 
$P_a=896$ s, $P_b=1603$ s, $P_c=2688$ s, $P_d=4062$ s, and $P_e=8123$ s, 
are indicated in Figure \ref{fig:rms}.
The five periods might be linear combinations of the two basic periods.
These appear to be related to each other: 
$P_b^{-1} \fallingdotseq P_c^{-1} + P_d^{-1}$,
$P_c = 3P_a$,
$2P_d \fallingdotseq 9P_a$,
and $P_e \fallingdotseq 2P_d$.
Thus, we regard the shortest period $P_a$ as $P_2$, which is a period independent from $P_1$.
This $P_2$ corresponds to the third strongest peak, $f_c$, in Figure \ref{fig:periodogram}. 
We also estimated the uncertainty of $P_2$ with the Monte Carlo technique as that of $P_1$.
We derived $P_2$ as \Ptwoerr~s with randomly resampled \Nmc lightcurves.

The model curves with $P_1$ of \Pone~s and $P_2$ of \Ptwo~s are shown in Figure \ref{fig:plc2d} with the observed lightcurves.
The RMS is calculated as 0.069.
The model and observed lightcurves are well overlapped each other,
which indicates that $P_1$ and $P_2$ are periods characterizing the rotation state of 2015 RN$_{35}$.
We note that $P_1$ and $P_2$ may not necessarily correspond to rotation and precession (or precession and rotation) periods, respectively.
Determination of the rotation and precession periods need detailed physical modeling, 
which is out of the scope of this paper.

\begin{figure*}[ht]
\plotone{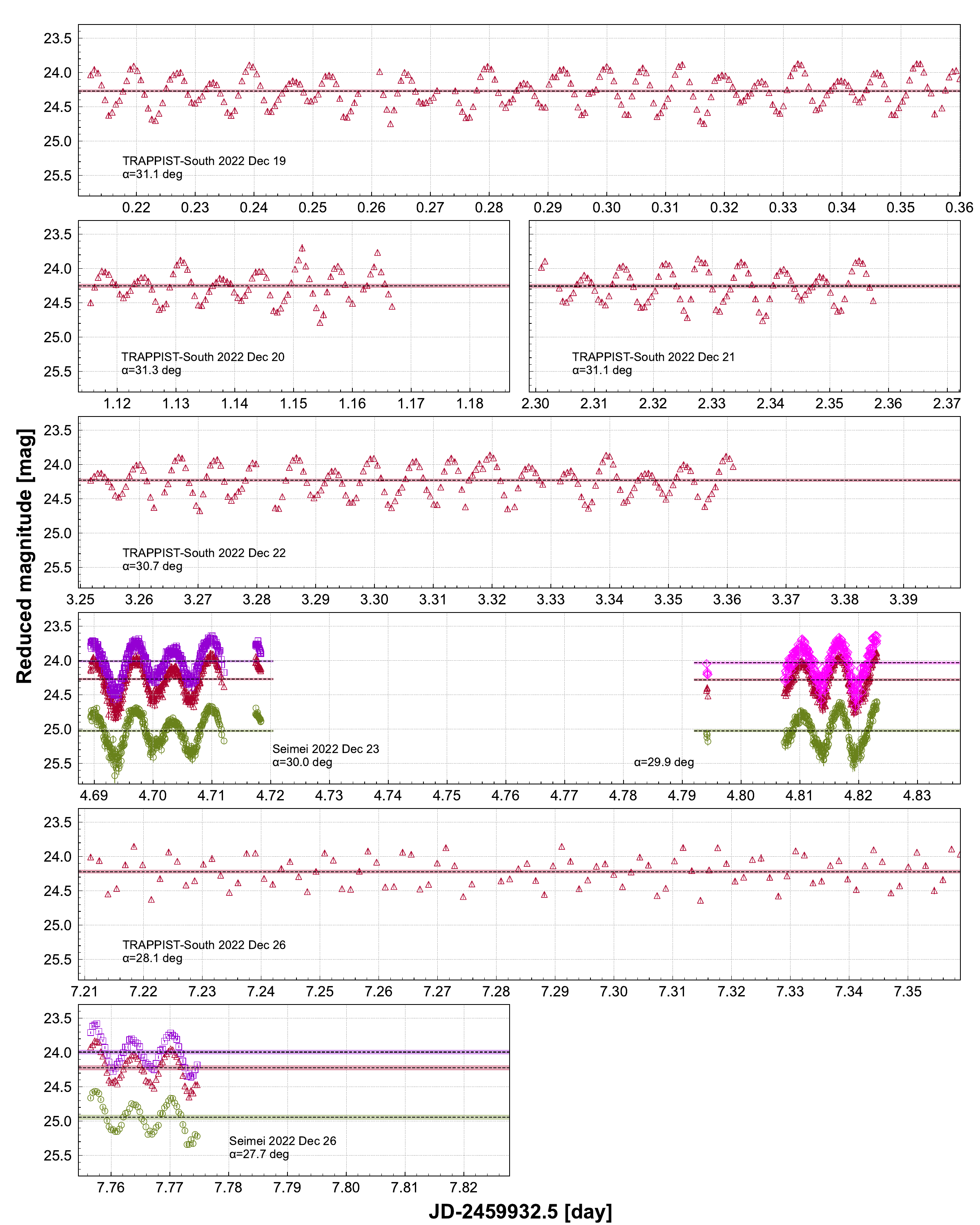}
\caption{Lightcurves of 2015 RN$_{35}$.
The reduced $g$, $r$, $i$, and $z$ bands magnitudes 
are presented as circles, triangles, squares, and diamonds, respectively.
Bars indicate the 1$\sigma$ uncertainties.
Error-weighted average of magnitude in each lightcurve is presented with a dashed line. 
Shaded areas indicate the standard errors of the weighted averaged magnitudes.
}\label{fig:lc}
\end{figure*}

\begin{figure}[ht]
\plotone{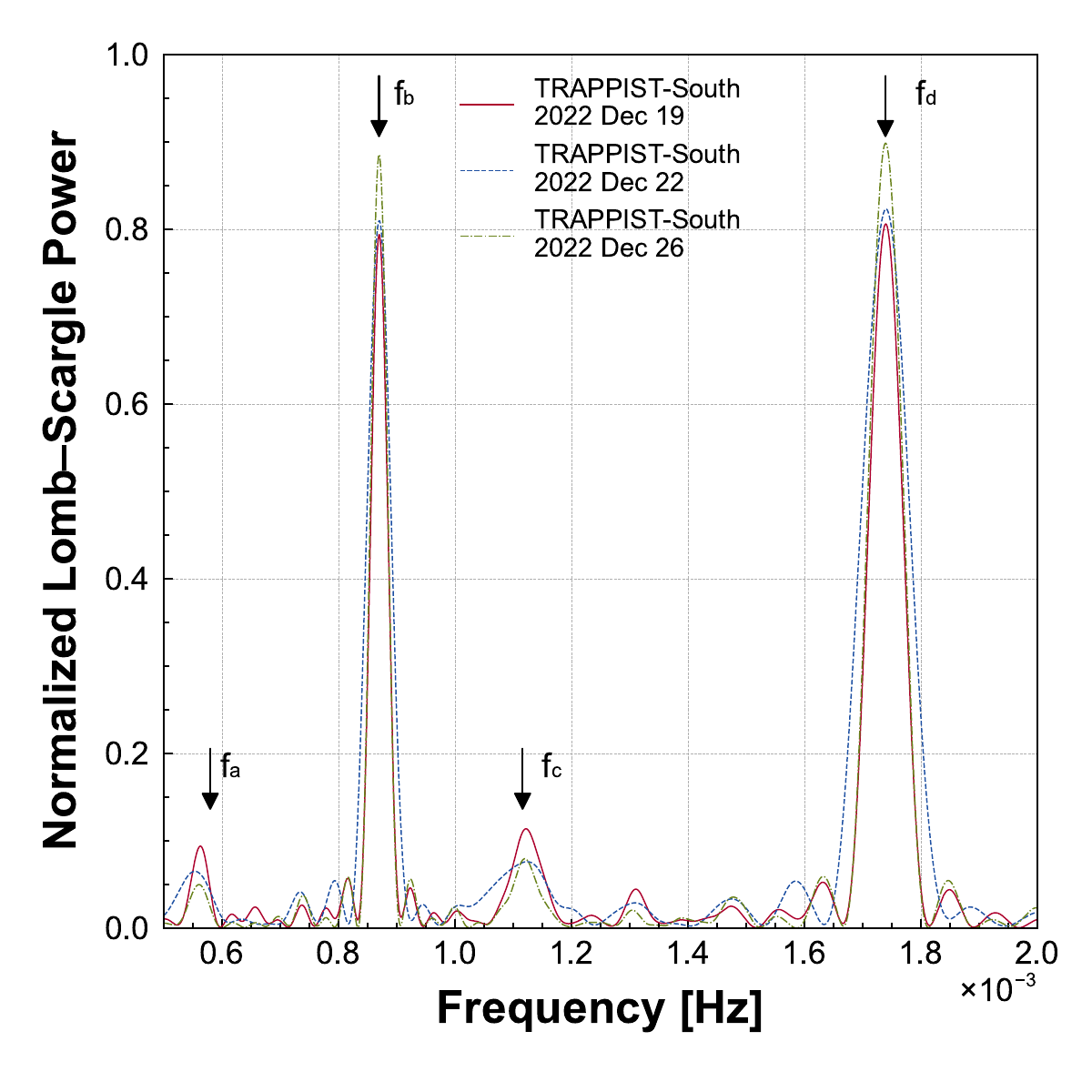}
\caption{
Lomb--Scargle periodogram of 2015 RN$_{35}$.
The number of harmonics is two.
The three periodograms were created using lightcurves obtained with the TRAPPIST-South telescope on
2022 December 19, 22, and 26.
Four peak frequencies in either case are indicated. 
}\label{fig:periodogram}
\end{figure}

\begin{figure*}[ht]
\plotone{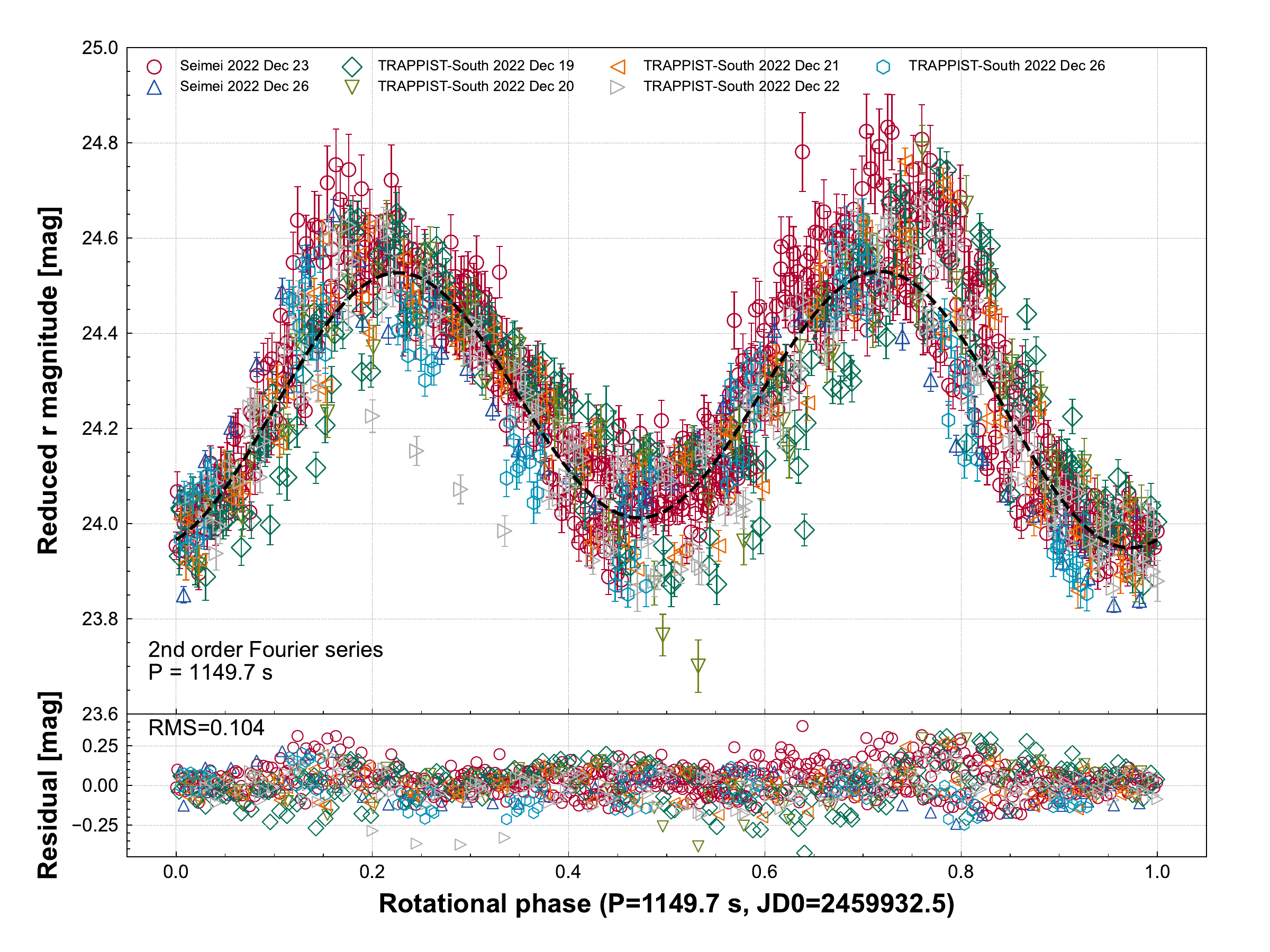}
\caption{Phased lightcurves of 2015 RN$_{35}$.
All lightcurves in reduced $r$ magnitude are folded with a period of \Pone~s.
Phase zero is set to \JDzero.
Bars indicate the 1$\sigma$ uncertainties of measurements.
Model curve fitted to lightcurves is shown by dashed line.
}\label{fig:plc}
\end{figure*}

\begin{figure*}[ht]
\plotone{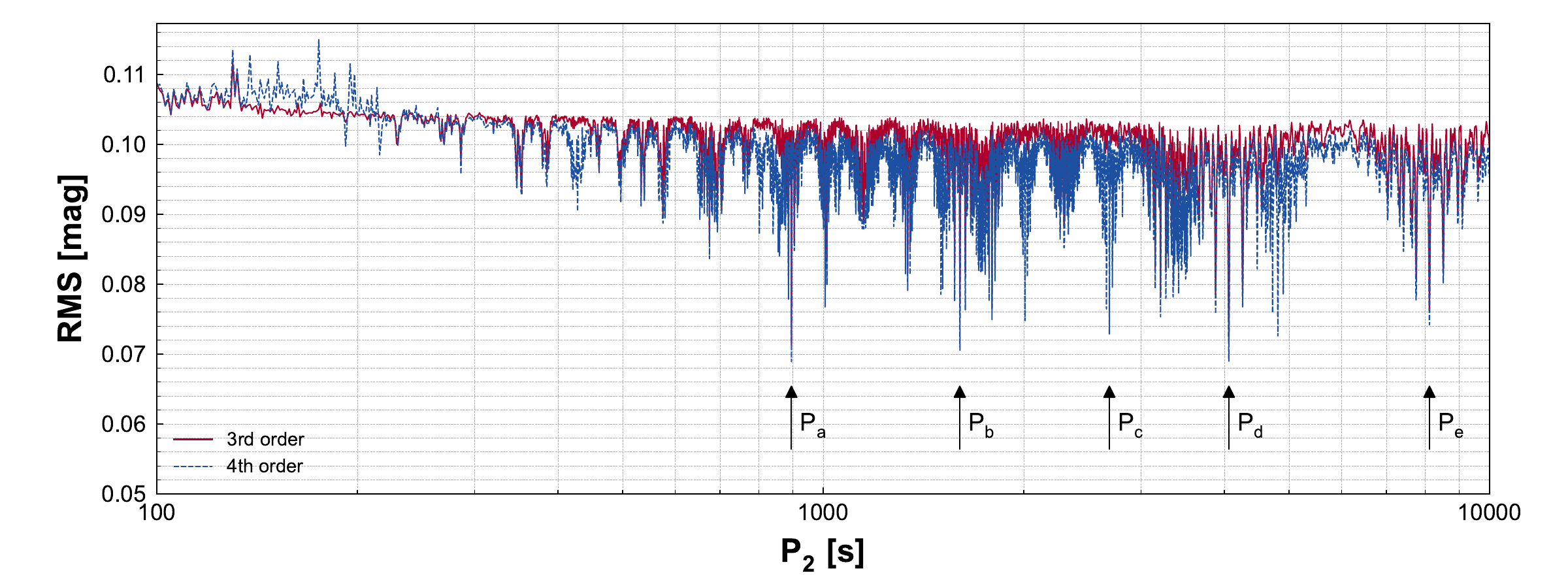}
\caption{
RMS residuals between observed and model lightcurves of 2015 RN$_{35}$ fixing $P_1$ of \Pone~s.
Residuals using two-dimensional Fourier series with $m$ of 3
and $m$ of 4 are shown by solid and dashed lines, respectively.
Five periods with smaller RMS in either case are indicated. 
}\label{fig:rms}
\end{figure*}

\begin{figure*}[ht]
\plotone{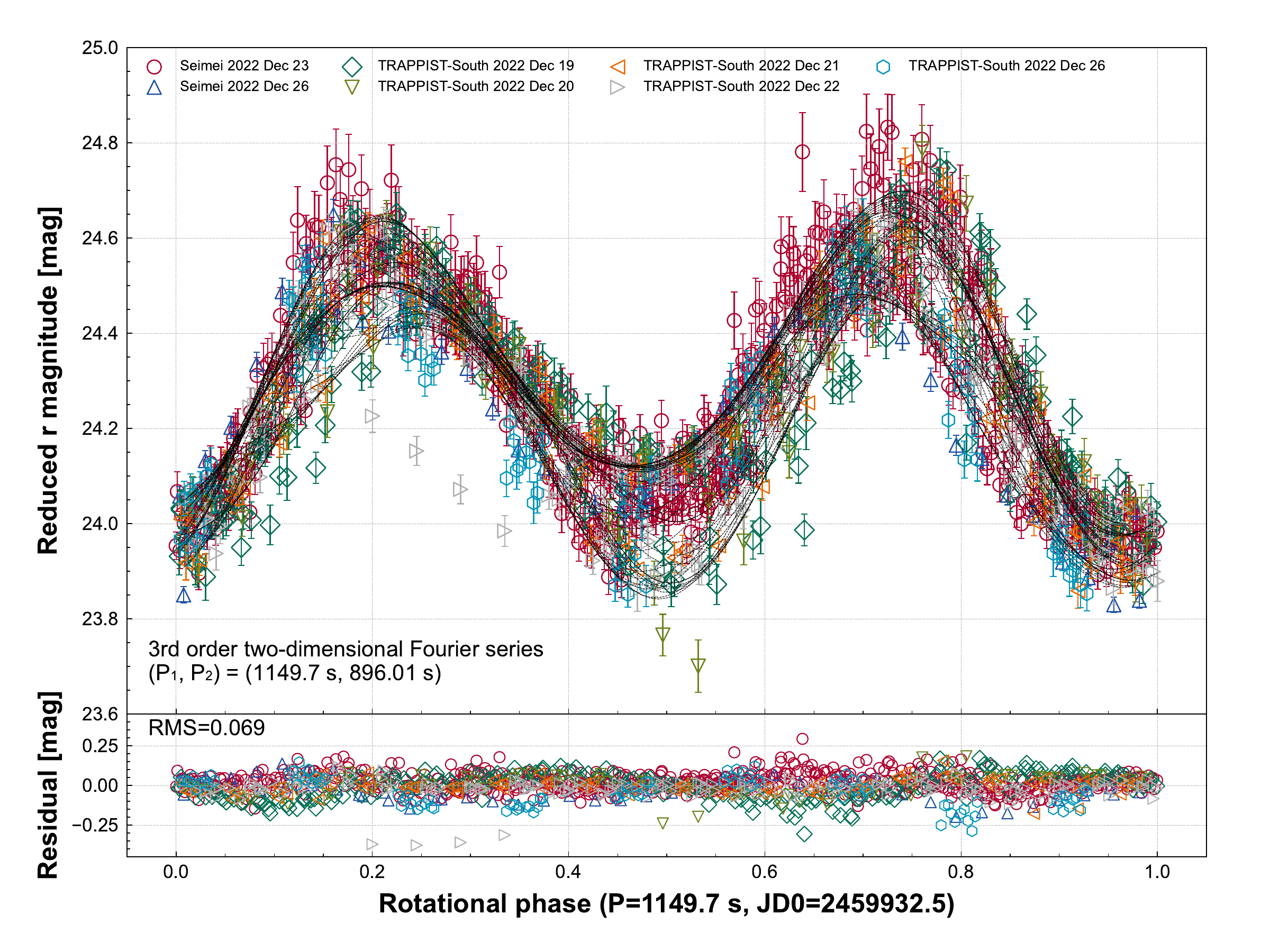}
\caption{
Phased lightcurves of 2015 RN$_{35}$ with two-dimensional Fourier series.
All lightcurves in reduced $r$ magnitude are folded with $P_1$ of \Pone~s and $P_2$ of \Ptwo~s.
Phase zero is set to \JDzero.
Bars indicate the 1$\sigma$ uncertainties.
Model curves fitted to lightcurves are shown by dashed lines.
}
\label{fig:plc2d}
\end{figure*}

\subsection{Colors and reflectance spectrum} \label{subsec:res_color}
The time-series of colors of 2015 RN$_{35}$ in 2022 December are shown in Figure \ref{fig:col}.
The error-weighted averages of those colors are derived as 
$g-r =$\gr, $r-i = $\ri, and $r-z = $\rz.
The systematic errors in color determination with TriCCS are considered as in \cite{Beniyama2023a}.
The error-weighted average colors corresponds each other within the measurement errors when we consider the results on 2023 January 
(see panel (b) of Figure \ref{fig:pc}).

The reflectance spectrum of 2015 RN$_{35}$ in Figure \ref{fig:ref} was calculated with 
the derived colors and the solar colors with the same method in \cite{Beniyama2023b}.
The reflectances at the central wavelength of the $r$, $i$, and $z$ bands, 
$R_r$, $R_i$, and $R_z$, were calculated as:
\begin{eqnarray}
    R_r &=& 10^{-0.4[(r-g)_{\mathrm{RN_{35}}}-(r-g)_\odot]}, \\
    R_i &=& 10^{-0.4[(i-g)_{\mathrm{RN_{35}}}-(i-g)_\odot]}, \\
    R_z &=& 10^{-0.4[(z-g)_{\mathrm{RN_{35}}}-(z-g)_\odot]}, 
\end{eqnarray}
where 
$(r-g)_\mathrm{RN_{35}}$, $(i-g)_\mathrm{RN_{35}}$, and $(z-g)_\mathrm{RN_{35}}$ 
are the colors of 2015 RN$_{35}$, 
whereas
$(r-g)_\odot$, $(i-g)_\odot$, and $(z-g)_\odot$ 
are the colors of the Sun in the Pan-STARRS system.
We referred to the absolute magnitude of the Sun in the Pan-STARRS system as
$g = 5.03$, $r=4.64$, $i=4.52$, and $z=4.51$ \citep{Willmer2018}.
We set the uncertainties of the magnitudes of the Sun as 0.02.

The reflectance spectra in Figure \ref{fig:ref} 
are normalized at the center of the $g$ band in the Pan-STARRS system, \gcenter \citep{Tonry2012}.
The horizontal bars in the 2015 RN$_{35}$'s spectrum indicate the filter bandwidths.
The reflectance spectra except 2015 RN$_{35}$ are originally normalized at 0.55~$\mu$m.
We renormalize those spectra at \gcenter as follows:
\begin{equation}
    R^{\prime}(\lambda) = \frac{R(\lambda)}{R(0.481\,\mu m)},
\end{equation}
where $R^{\prime}(\lambda)$ is a renormalized reflectance at a wavelength of $\lambda$, 
$R(\lambda)$ is an original reflectance at a wavelength of $\lambda$,
and $R(0.481\,\mu m)$ is an original reflectance at a wavelength of \gcenter.

\begin{figure*}[ht]
\plotone{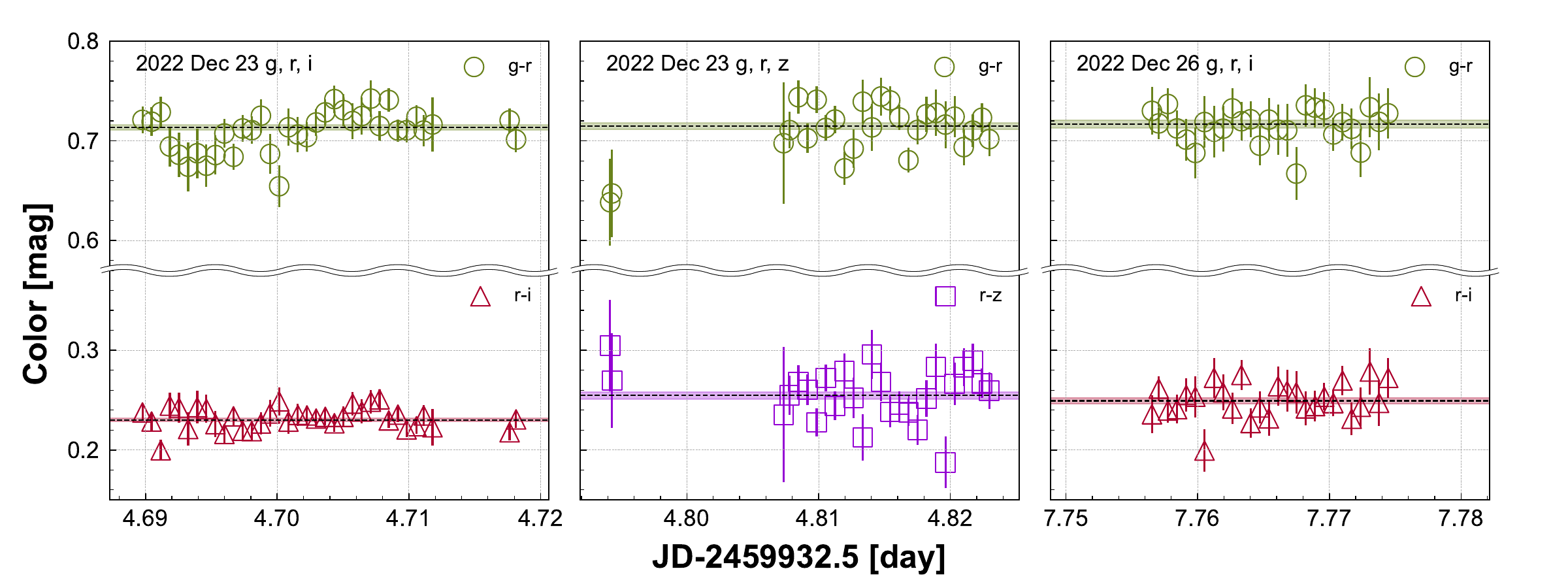}
\caption{
Time-series colors of 2015 RN$_{35}$.
The $g-r$, $r-i$, and $i-z$ colors are shown by circles, triangles, and squares, respectively.
Binning of 60~s are performed for all colors.
Bars indicate the 1$\sigma$ uncertainties.
Weighted mean and its error are indicated by dashed lines and shaded areas, respectively.
}\label{fig:col}
\end{figure*}

\begin{figure}[ht]
\plotone{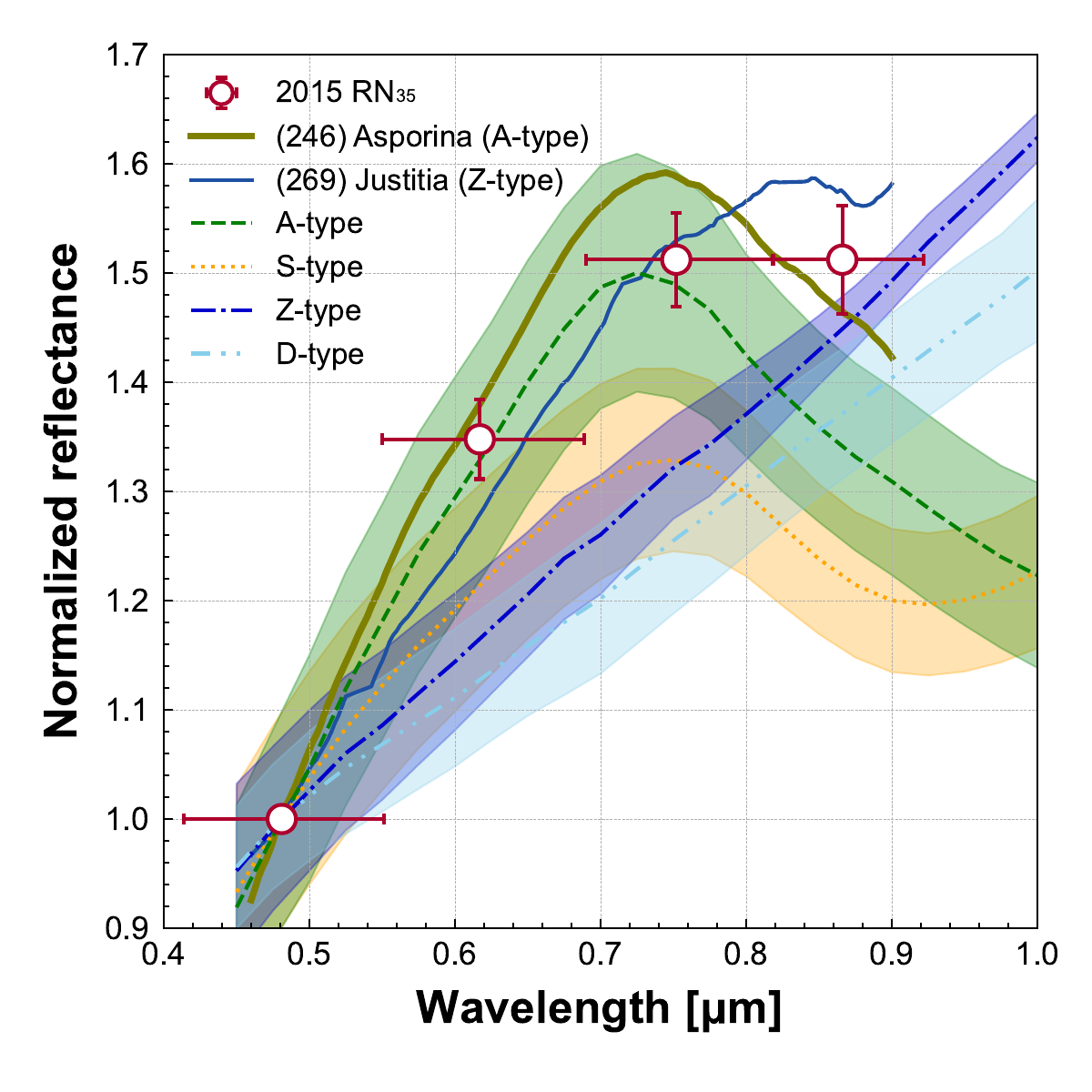}
\caption{
Reflectance spectrum of 2015 RN$_{35}$ (circles).
Vertical bars indicate the 1$\sigma$ uncertainties.
Horizontal bars indicate the filter bandwidths.
Template spectra of A- (dashed line), S- (dotted line), 
Z- (dot-dashed line),
and D-types (dot-dot-dashed line)
are shown \citep{Mahlke2022}.
Shaded areas indicate the standard deviations of the template spectra.
Visible spectra of the A-type asteroid (246) Asporina (thick solid line)
and the Z-type asteroid (269) Justitia (thin solid line)
are shown.
The reflectance spectra are normalized at \gcenter.
}\label{fig:ref}
\end{figure}

\subsection{Phase curves} \label{subsec:res_pc}
We observed 2015 RN$_{35}$ across a wide range of phase angles from 2$\degdeg$ to 30$\degdeg$, 
which provides us a well-sampled phase curve as shown in Figure \ref{fig:pc}.
We converted the $g$ and $r$ band magnitudes in the Pan-STARRS system to the V band magnitude in the Johnson system using the equations in \citet{Tonry2012}.
We stacked 20 images obtained in 2023 January to 
make a decent detection of \RN for photometry as shown in the upper panels of Figure \ref{fig:cutout}.
The total integration time, 1200~s, is compatible to one of the characteristic period of \RN, $P_1$ of \Pone~s.
Thus, the rotation effects have been corrected in the phase curves of \RN.

We derived an absolute magnitude in the V band, $H$, and slope parameters, $G_1$ and $G_2$, 
with the $H$-$G_1$-$G_2$ model \citep{Muinonen2010}:
\begin{align}
    V_\mathrm{red}(\alpha) = &H - 2.5 \log_{10}\\ \nonumber 
                             &{[G_1\Phi_1(\alpha) + G_2\Phi_2(\alpha) + (1-G_1-G_2)\Phi_3(\alpha)]},
\end{align}
where $V_\mathrm{red}(\alpha)$ is a reduced magnitude in the V band at a phase angle of $\alpha$.
The $\Phi_1$, $\Phi_2$, and $\Phi_3$ are
phase functions written as follows:
\begin{eqnarray}
    \Phi_1(\alpha) &=& 1 - \frac{6\alpha}{\pi},\\
    \Phi_2(\alpha) &=& 1 - \frac{9\alpha}{5\pi},\\
    \Phi_3(\alpha) &=& \exp\left(-4\pi \tan^{2/3}{\frac{\alpha}{2}}\right).
\end{eqnarray}
The uncertainty of $H$, $G_1$, and $G_2$ were estimated with the Monte Carlo technique.
We made \Nmc phase curves by randomly resampling the data assuming each observed data 
follows a normal distribution whose standard deviation is a standard error of weighted mean magnitude.
We derived $H$ of \HV, $G_1$ of \Gone, and $G_2$ of \Gtwo.
The absolute magnitude and slope parameters in the $g$, $r$, and $i$ bands are also derived in Figure \ref{fig:pc} for convenience.

\begin{figure*}[ht]
\plotone{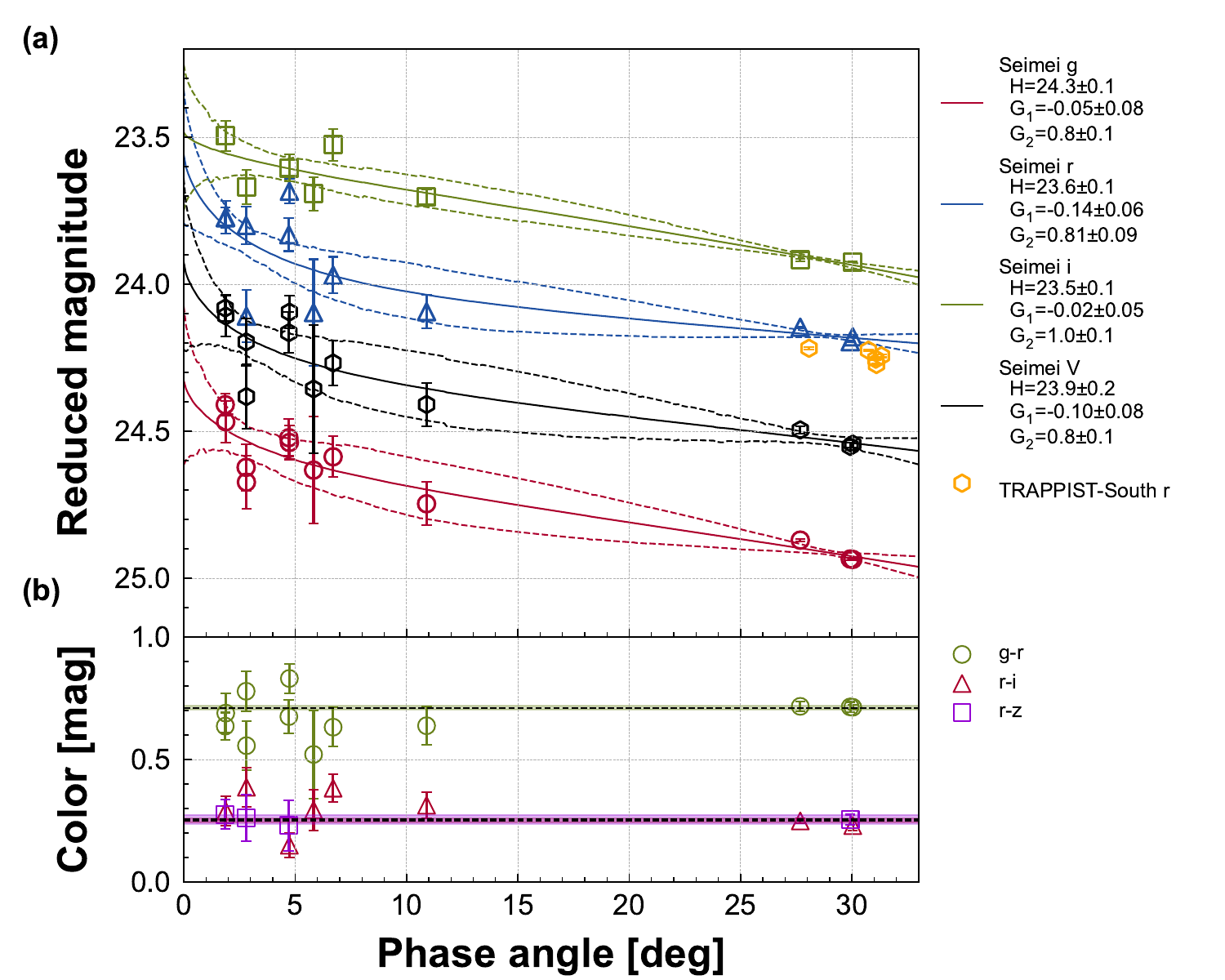}
\caption{
    Phase angle dependence of magnitude and colors of 2015 RN$_{35}$.
    (a)
    Mean reduced $g$, $r$, $i$, $z$, and V band 
    magnitudes are presented as circles, triangles, squares, diamonds, and hexagons, respectively.
    Bars indicate the 1$\sigma$ uncertainties.
    Median (50th percentile) of fitting model curves with the $H$-$G_1$-$G_2$ model are presented by solid lines.
    Uncertainty envelopes representing the 95~\% highest density inverval (HDI) values are shown by dashed lines.
    (b) 
    Weighted mean $g-r$, $r-i$, and $r-z$ colors and their errors in each day are presented as circles, triangles, and squares, respectively.
    Bars indicate the 1$\sigma$ uncertainties.
    Global weighted mean colors and their standard errors are indicated by dashed lines and shaded areas, respectively
    although they are small and hard to see due to scale effects.
    }
\label{fig:pc}
\end{figure*}

\section{Discussion} \label{sec:discussion}
\subsection{Possible classification of \RN} \label{subsec:possible}
The visible spectrum of \RN suggests that \RN is a very red object (VRO) in the near-Earth region. 
We compare the spectrum with the class templates from \cite{Mahlke2022} in Figure \ref{fig:ref}.
We also show the spectra of the A-type MBA (246) Asporina and the VRO in the main belt (269) Justitia.
These two spectra were obtained with SpeX \citep{Rayner2003} on the NASA Infrared Telescope Facility (IRTF).
We obtained the two spectra via the M4AST online tool \citep[Modeling for Asteroids;][]{Popescu2012}.
Justitia is known to have a very red slope like the trans-Neptunian objects \citep[TNOs;][]{Hasegawa2021b},
and is classified as Z-type in the latest Mahlke taxonomy \citep{Mahlke2022}.
We note that the spectrum of Justitia seems to be out of the range of the Z-type template in the visible wavelength in Figure \ref{fig:ref}.
This is because Justitia is classified as Z-type in \cite{Mahlke2022} using both visible and near-infrared spectra.
Thus, Justitia is a bit redder than the typical Z-types.
The Z-types have featureless and extremely redder spectra than the D-types.

We evaluate the goodness-of-fit between the spectrum of \RN and templates
using the following quantity:
\begin{equation}
    \delta^2 = \frac{1}{N}\sum_{j} (R_{\mathrm{obs}, j}-R_{\mathrm{model}, j})^2,
\end{equation}
where $N$ is the number of reflectance values,
$R_{\mathrm{obs}, j}$ is a reflectance of \RN at $j$th wavelength,
and $R_{\mathrm{model}, j}$ is a reflectance of a template spectrum at the wavelength.
We found that the spectrum of 2015 RN$_{35}$ seems like those of A-types ($\delta^2=0.008$) and Z-types ($\delta^2=0.018$), whereas the spectrum does not fit well with S-types ($\delta^2=0.034$) and D-types ($\delta^2=0.033$).
We note that only visible colors are often not enough to determine the spectral types of asteroids.
For instance, half of all objects classified as A-types based on spectra in the visible wavelength
are not A-types in the near-infrared \citep{DeMeo2019}.

We classified \RN as an A- or Z-type in this study,
where both types represent rare populations \citep{Mahlke2022}.
The A-type asteroids are olivine-rich asteroids and have similar spectra to those of silicate mineral olivine
and are thought to be a piece of differentiated planetesimal \citep{DeMeo2019},
while other studies propose that some A-types may originate from the mantle of the Mars \citep{Polishook2017a}.
Thus, the A-types may have an important role to investigate formation of terrestial planets.
Recently, the two VROs, (203) Pompeja and (269) Justitia, are discovered in \cite{Hasegawa2021b}.
\cite{BourdelledeMicas2022} discovered the VRO (732) Tjilaki in the main belt.
These VROs are classified as Z-types in the latest Mahlke taxonomy \citep{Mahlke2022}.
In total, 23 asteroids including one NEA, (141670) 2002 JS$_{100}$, are classified as Z-type in \cite{Mahlke2022}.
The Z-types might have primitive organic materials on the surface as D-types \citep{Barucci2018}.
Justitia is selected as the rendezvous target of the Emirates Mission to Explore the Asteroid Belt \citep{Alhameli2023}.

It is known that the slope parameters $G_1$ and $G_2$ have a tight correlation with the geometric albedo \citep{Muinonen2010, Shevchenko2016}.
We show typical $G_1$ and $G_2$ values of A-, E-, S-, C- and D-types in Figure \ref{fig:G1G2} \citep{Shevchenko2016, Mahlke2021}.
We also plot the $G_1$ and $G_2$ of the A-type MBA (246) Asporina derived using 
the sparse photometric observations from Gaia Data Release 2 \citep{Martikainen2021}.
The smaller $G_1$ and larger $G_2$, by definition, mean that the slope of the phase curve is shallower.
The slopes of high-albedo asteroid are shallower since
the contribution of the shadow-hiding effect decreases as albedo increases \citep[e.g.;][]{Belskaya2000}, whereas those of low-albedo asteroids are steeper on the contrary.
Thus, the small $G_1$ and large $G_2$ of \RN are indicative of a high geometric albedo.

The $G_1$ and $G_2$ of \RN seems a bit far from the typical values of A-types on Figure \ref{fig:G1G2}.
But, the typical values are slightly different from each other of about 0.1--0.2 on $G_1$--$G_2$ plane depending on the references.
Thus, the discrepancy between the slope parameters of \RN and typical values does not necessarily indicate that \RN is an outlier.
Therefore we concluded that \RN is an A-type asteroid in conjunction with the colors and slope parameters in the visible wavelength.
We demonstrated that only photometry in the visible wavelength is 
sufficient to determine the spectral type of asteroids, if it is across a wide range of phase angles.

Finally, we discuss other interpretations of the shallow phase slope of \RN.
The environments such as self-gravity and rotation period are different between small and large asteroids.
Small asteroids may have different surface properties compared with large asteroids.
\cite{Terai2013} observed a tiny L-type NEA (367943) Duende (a.k.a., 2012 DA$_{14}$) 
across a wide phase angle range from 19$\degdeg$ to 42$\degdeg$. 
They derived the slope parameter in the $H$-$G$ model, $G$, as $0.44^{+0.06}_{-0.08}$,
which is larger than the typical value of L-types.
The tiny asteroid Duende has a shallow slope in the phase curve.
They interpreted the large slope parameter or shallow slope with the difference of
surface environment due to the lack of the fine regolith or high geometric albedo.
Small gravity on tiny asteroids might lead to the lack of the fine regolith on its surface.
The shadow-hiding effect is weak when the fine regolith is deficient and the slope of the phase curve would be shallow.
As for \RN, in addition to the small gravity, the fast rotation of about 20~minutes also supports the hypothesis of the lack of fine regolith. 
Other recent studies showed that there is almost no correlation between albedo and slope parameters 
using the phase curves obtained in the frame work of the IMPACTON project and the ATLAS survey \citep{Rondon2019, Arcoverde2023}.
They interpreted this is due to the difference of diameters between NEAs and MBAs.
The trend is not clear since the number of samples are limited due to the observational difficulties.
A comprehensive research of phase parameters are desired to reach a conclusion.

Various observations such as near-infrared spectroscopy and polarimetry are 
crucial in forthcoming approaches of \RN to put an end to the spectral type.
The next two opportunities are in 2031 September and 2056 December,
where \RN will be brightened up to 21~mag and 16~mag in the V band, respectively.

\begin{figure*}[ht]
\plotone{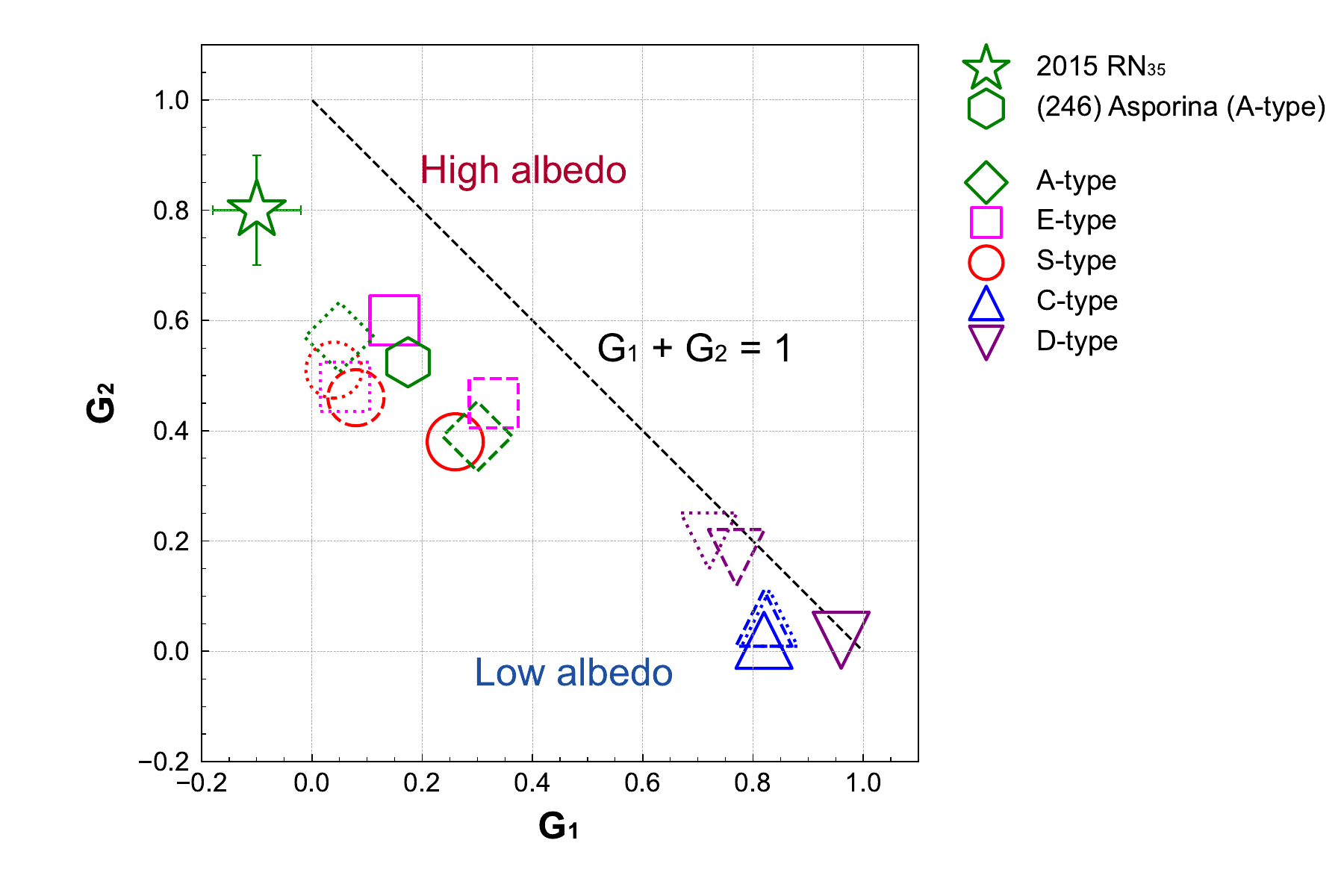}
\caption{
    $G_1$ and $G_2$ of \RN in the V band (star).
    Bars indicate the 1$\sigma$ uncertainties.
    $G_1$ and $G_2$ of the A-type asteroid (246) Asporina is shown by hexagon \citep{Martikainen2021}.
    Typical $G_1$ and $G_2$ values of A-, E-, S-, C-, and D-types are shown by 
    diamonds, squares, circles, triangles, and inverted triangles, respectively.
    Markers are enclosed by solid lines \citep{Shevchenko2016}, dashed lines \citep[cyan band;][]{Mahlke2021}, and dotted lines \citep[orange band;][]{Mahlke2021}.
    Isochrone for $G_1 + G_2 = 1$  
    is plotted by the dashed line for convenience.
}\label{fig:G1G2}
\end{figure*}

\subsection{Diameter estimation of tiny asteroids} \label{subsec:diam}
Estimation of asteroid sizes is important not only to evaluate a risk of impact to the Earth but also to plan exploration missions.
However, small NEAs are often observed only at a few apparitions at 
relatively large phase angles compared to MBAs and TNOs.
Thus, the absolute magnitude of the asteroid is often not well constrained.
The absolute magnitude could be uncertain by $\sim0.3$ depending on whether the opposition surge exists \citep[see, e.g.;][]{Belskaya2003}.
\cite{Juric2002} and \cite{Pravec2012} independently estimated
that there is a systematic uncertainty of $H$ of about 0.4.
In addition to the uncertainty of $H$, the geometric albedo is not well estimated for small bodies 
since observing opportunities are limited to only a short period, as this study.

We estimated the absolute magnitude of \RN with high accuracy as \HV 
through observations across a wide range of phase angles down to 2$\degdeg$.
The surface colors as well as the slope of the phase curve indicate that 
\RN is a very red asteroid, probably classified as an A-type asteroid.
The typical geometric albedo of A-types is 
estimated as $0.282\pm0.101$ and $0.28\pm0.09$ in \cite{Usui2013} and \cite{DeMeo2019}, respectively.
We assume the geometric albedo of \RN as $p_\mathrm{V}$ of $0.28\pm0.10$.
The diameter of an asteroid can be estimated with $H$ and $p_\mathrm{V}$
using the following equation \citep{Fowler1992, Pravec2007}:
\begin{equation}
    D=\frac{1329}{\sqrt{p_\mathrm{V}}}\times10^{-H/5}. \label{eq:D}
\end{equation}
The diameter of 2015 RN$_{35}$ is estimated to be \diam m.
We updated absolute magnitude of \RN by about 0.7 compared to $H$ of \HJPL in JPL SBDB.
Our study demonstrated that it is crucial to observe the asteroid in multibands at 
multi-epochs including where the phase angles are low to derive $H$ and $D$ accurately.
Observations at very low phase angles are not possible for all NEAs.
Detailed planning of observations is crucial for the diameter estimation of tiny asteroids.

\subsection{Mission accessibility} \label{subsec:mission}
One of the important parameters to plan the spacecraft mission is the delta-v ($\Delta v$),
which is the required impulse per unit of spacecraft mass to change the status of the spacecraft.
We refer to the total $\Delta v$ \footnote{\url{https://cneos.jpl.nasa.gov/nhats/}} as the sum of 
(i) the maneuver required to depart a notional 400~km altitude circular Earth parking orbit, 
(ii) the $\Delta v$ required to match the NEA's velocity at arrival, 
(iii) the $\Delta v$ required to depart the NEA, 
and (iv) the $\Delta v$ (if any) required to control atmospheric entry speed at Earth return.
We queried the $\Delta v$ of NEAs for the Near-Earth Object Human Space Flight Accessible Targets Study \citep[NHATS;][]{Abell2012}
\footnote{\url{https://ssd-api.jpl.nasa.gov/doc/nhats.html}, last accessed 2023 August 10.}.
The $\Delta v$ of \RN is estimated as 11.801~km\,s$^{-1}$ in the launch window between 2030 and 2035,
which is smaller than the limit of NHATS, 12~km\,s$^{-1}$.
In terms of the engineering aspect, \RN is a good candidate for a future spacecraft mission.
From a scientific point of view, \RN, either it is an A- or a Z-type NEA, is a great candidate for a future mission. 
Specifically, there are no planned future spacecraft missions to A-type asteroids.

\section{Conclusions} \label{sec:conclusions}
We conducted multicolor photometry of the tiny NEA 2015 RN$_{35}$ over 17 nights in 2022 December and 2023 January.
We observed 2015 RN$_{35}$ across a wide range of phase angles from 
2$^{\circ}$ to 30$^{\circ}$ in the $g$, $r$, $i$, and $z$ bands in the Pan-STARRS system.
We found that 2015 RN$_{35}$ is in a non-principal axis spin state
with two characteristic periods of \Poneerr~s and \Ptwoerr~s.
The visible spectrum of 2015 RN$_{35}$ is as red as (269) Justitia, one of the VROs in the main belt,
which indicates that 2015 RN$_{35}$ can be classified as an A- or Z-type asteroid.
Together with the shallow slope of the phase curve, we suppose 2015 RN$_{35}$ is a high-albedo A-type asteroid.

Various observations such as near-infrared spectroscopy and polarimetry are encouraged during the forthcoming approaches of \RN.
The next opportunity is in September 2031, where \RN will be brightened up to 21~mag.
Though additional follow-up observations are required to reach a conclusion,
2015 RN$_{35}$ is a possible mission accessible A-type NEA with 
small $\Delta v$ of 11.801~km\,s$^{-1}$ in a launch window between 2030 and 2035.

\begin{acknowledgments}
We sincerely thank Seitaro Urakawa, Miho Kawabata, 
Qiliang Fang, Keisuke Isogai, Hideyuki Izumiura, Kentaro Tanimoto, and Ren Ikeya for observing assistance.
Special thanks to Patricio Rojo for attempting to observe \RN in Chile.
We are grateful to Benoit Carry and Eric MacLennan 
for the discussions about the latest taxonomy and phase function, respectively.
J.B. also acknowledge the contribution of Sean Marshall for improving the quality of the manuscript.
We are grateful to the anonymous referee for constructive comments during the review process.
J.B. would like to express his gratitude to the Iwadare Scholarship Foundation, 
the Public Trust Iwai Hisao Memorial Tokyo Scholarship Fund,
the JEES Mitsubishi Corporation Science Technology Student Scholarship, and Iue Memorial Foundation for the grants.
C.A. acknowledges support from the ANR ORIGINS (ANR-18-CE31-0014).
This research is partially supported by the Optical and Infrared Synergetic Telescopes for Education and Research (OISTER) program funded by the MEXT of Japan.
This work is supported in part by the JST SPRING, grant No. JPMJSP2108, and the UTEC UTokyo Scholarship.
This work has been supported by the Japan Society for the Promotion of Science (JSPS) 
grants-in-aid for Scientific Research (KAKENHI) grant Nos. 21H04491 and 23KJ0640.
The authors thank the TriCCS developer team 
(which has been supported by the JSPS KAKENHI grant Nos. JP18H05223, 
JP20H00174, and JP20H04736, and by NAOJ Joint Development Research).
This research is based on data collected by the TRAPPIST-South telescope at the ESO La Silla Observatory. 
TRAPPIST is funded by the Belgian Fund for Scientific Research (Fond National de la Recherche Scientifique, FNRS) 
under the grant PDR T.0120.21. E.J. is a F.R.S.-FNRS Senior Research Associate.
The Pan-STARRS1 Surveys (PS1) and the PS1 public science archive have been made 
possible through contributions by the Institute for Astronomy, 
the University of Hawaii, the Pan-STARRS Project Office, the Max-Planck Society 
and its participating institutes, the Max Planck Institute for Astronomy, 
Heidelberg and the Max Planck Institute for Extraterrestrial Physics, 
Garching, The Johns Hopkins University, Durham University, the University of Edinburgh, 
the Queen's University Belfast, the Harvard-Smithsonian Center for Astrophysics, 
the Las Cumbres Observatory Global Telescope Network Incorporated, 
the National Central University of Taiwan, the Space Telescope Science Institute, 
the National Aeronautics and Space Administration under grant No. NNX08AR22G 
issued through the Planetary Science Division of the NASA Science Mission Directorate, 
the National Science Foundation grant No. AST-1238877, the University of Maryland, 
Eotvos Lorand University (ELTE), the Los Alamos National Laboratory, 
and the Gordon and Betty Moore Foundation.
Part of the data utilized in this publication were obtained and made available by 
the MIT-UH-IRTF Joint Campaign for NEO Reconnaissance. 
The IRTF is operated by the University of Hawaii under Cooperative Agreement 
no. NCC 5-538 with the National Aeronautics and Space Administration, Office of Space Science, 
Planetary Astronomy Program. 
The MIT component of this work is supported by NASA grant 09-NEOO009-0001, 
and by the National Science Foundation under Grants Nos. 0506716 and 0907766.
\end{acknowledgments}

\vspace{5mm}
\facilities{Seimei (TriCCS), TRAPPIST}

\software{
NumPy \citep{Oliphant2015, Harris2020},
pandas \citep{Reback2021},
SciPy \citep{Virtanen2020},
AstroPy \citep{Astropy2013, Astropy2018},
Astro-SCRAPPY \citep{McCully2018},
astroquery \citep{Ginsburg2019}, 
Matplotlib \citep{Hunter2007},
Source Extractor \citep{Bertin1996},
SEP \citep{Barbary2017},
astrometry.net \citep{Lang2010},
PHOTOMETRYPIPELINE \citep{Mommert2017}
}

\newcommand{\noopsort}[1]{} \newcommand{\singleletter}[1]{#1}

\bibliographystyle{aasjournal}

\begin{thebibliography}{}
\expandafter\ifx\csname natexlab\endcsname\relax\def\natexlab#1{#1}\fi
\providecommand{\url}[1]{\href{#1}{#1}}
\providecommand{\dodoi}[1]{doi:~\href{http://doi.org/#1}{\nolinkurl{#1}}}
\providecommand{\doeprint}[1]{\href{http://ascl.net/#1}{\nolinkurl{http://ascl.net/#1}}}
\providecommand{\doarXiv}[1]{\href{https://arxiv.org/abs/#1}{\nolinkurl{https://arxiv.org/abs/#1}}}

\bibitem[{{Abell} {et~al.}(2012){Abell}, {Barbee}, {Mink}, {Adamo},
  {Alberding}, {Mazanek}, {Johnson}, {Yeomans}, {Chodas}, {Chamberlin},
  {Benner}, {Drake}, \& {Friedensen}}]{Abell2012}
{Abell}, P.~A., {Barbee}, B.~W., {Mink}, R.~G., {et~al.} 2012, in 43rd Annual
  Lunar and Planetary Science Conference, Lunar and Planetary Science
  Conference, 2842

\bibitem[{{Alhameli} {et~al.}(2023){Alhameli}, {Parker}, {Caudill}, {Baskar},
  {Rosen}, {Koehler}, {Chikine}, \& {Imler}}]{Alhameli2023}
{Alhameli}, F.~S., {Parker}, J.~S., {Caudill}, M.~T., {et~al.} 2023, in
  Asteroids, Comets, and Meteors: ACM 2023

\bibitem[{{Arcoverde} {et~al.}(2023){Arcoverde}, {Rond{\'o}n}, {Monteiro},
  {Pereira}, {Ieva}, {Michtchenko}, {Evangelista-Santana}, {Michimani},
  {Mesquita}, {Corr{\^e}a}, {Dotto}, {Giunta}, {Di Paola}, {Medeiros},
  {Carvano}, {Rodrigues}, \& {Lazzaro}}]{Arcoverde2023}
{Arcoverde}, P., {Rond{\'o}n}, E., {Monteiro}, F., {et~al.} 2023, \mnras, 523,
  739, \dodoi{10.1093/mnras/stad1486}

\bibitem[{{Astropy Collaboration} {et~al.}(2013){Astropy Collaboration},
  {Robitaille}, {Tollerud}, {Greenfield}, {Droettboom}, {Bray}, {Aldcroft},
  {Davis}, {Ginsburg}, {Price-Whelan}, {Kerzendorf}, {Conley}, {Crighton},
  {Barbary}, {Muna}, {Ferguson}, {Grollier}, {Parikh}, {Nair}, {Unther},
  {Deil}, {Woillez}, {Conseil}, {Kramer}, {Turner}, {Singer}, {Fox}, {Weaver},
  {Zabalza}, {Edwards}, {Azalee Bostroem}, {Burke}, {Casey}, {Crawford},
  {Dencheva}, {Ely}, {Jenness}, {Labrie}, {Lim}, {Pierfederici}, {Pontzen},
  {Ptak}, {Refsdal}, {Servillat}, \& {Streicher}}]{Astropy2013}
{Astropy Collaboration}, {Robitaille}, T.~P., {Tollerud}, E.~J., {et~al.} 2013,
  \aap, 558, A33, \dodoi{10.1051/0004-6361/201322068}

\bibitem[{{Astropy Collaboration} {et~al.}(2018){Astropy Collaboration},
  {Price-Whelan}, {Sip{\H{o}}cz}, {G{\"u}nther}, {Lim}, {Crawford}, {Conseil},
  {Shupe}, {Craig}, {Dencheva}, {Ginsburg}, {VanderPlas}, {Bradley},
  {P{\'e}rez-Su{\'a}rez}, {de Val-Borro}, {Aldcroft}, {Cruz}, {Robitaille},
  {Tollerud}, {Ardelean}, {Babej}, {Bach}, {Bachetti}, {Bakanov}, {Bamford},
  {Barentsen}, {Barmby}, {Baumbach}, {Berry}, {Biscani}, {Boquien}, {Bostroem},
  {Bouma}, {Brammer}, {Bray}, {Breytenbach}, {Buddelmeijer}, {Burke},
  {Calderone}, {Cano Rodr{\'\i}guez}, {Cara}, {Cardoso}, {Cheedella}, {Copin},
  {Corrales}, {Crichton}, {D'Avella}, {Deil}, {Depagne}, {Dietrich}, {Donath},
  {Droettboom}, {Earl}, {Erben}, {Fabbro}, {Ferreira}, {Finethy}, {Fox},
  {Garrison}, {Gibbons}, {Goldstein}, {Gommers}, {Greco}, {Greenfield},
  {Groener}, {Grollier}, {Hagen}, {Hirst}, {Homeier}, {Horton}, {Hosseinzadeh},
  {Hu}, {Hunkeler}, {Ivezi{\'c}}, {Jain}, {Jenness}, {Kanarek}, {Kendrew},
  {Kern}, {Kerzendorf}, {Khvalko}, {King}, {Kirkby}, {Kulkarni}, {Kumar},
  {Lee}, {Lenz}, {Littlefair}, {Ma}, {Macleod}, {Mastropietro}, {McCully},
  {Montagnac}, {Morris}, {Mueller}, {Mumford}, {Muna}, {Murphy}, {Nelson},
  {Nguyen}, {Ninan}, {N{\"o}the}, {Ogaz}, {Oh}, {Parejko}, {Parley}, {Pascual},
  {Patil}, {Patil}, {Plunkett}, {Prochaska}, {Rastogi}, {Reddy Janga},
  {Sabater}, {Sakurikar}, {Seifert}, {Sherbert}, {Sherwood-Taylor}, {Shih},
  {Sick}, {Silbiger}, {Singanamalla}, {Singer}, {Sladen}, {Sooley},
  {Sornarajah}, {Streicher}, {Teuben}, {Thomas}, {Tremblay}, {Turner},
  {Terr{\'o}n}, {van Kerkwijk}, {de la Vega}, {Watkins}, {Weaver}, {Whitmore},
  {Woillez}, {Zabalza}, \& {Astropy Contributors}}]{Astropy2018}
{Astropy Collaboration}, {Price-Whelan}, A.~M., {Sip{\H{o}}cz}, B.~M., {et~al.}
  2018, \aj, 156, 123, \dodoi{10.3847/1538-3881/aabc4f}

\bibitem[{{Avdellidou} {et~al.}(2020){Avdellidou}, {Di Donna}, {Schultz},
  {Harthong}, {Price}, {Peyroux}, {Britt}, {Cole}, \&
  {Delbo'}}]{Avdellidou2020}
{Avdellidou}, C., {Di Donna}, A., {Schultz}, C., {et~al.} 2020, \icarus, 341,
  113648, \dodoi{10.1016/j.icarus.2020.113648}

\bibitem[{{Barbary} {et~al.}(2017){Barbary}, {Boone}, {Craig}, {Deil}, \&
  {Rose}}]{Barbary2017}
{Barbary}, K., {Boone}, K., {Craig}, M., {Deil}, C., \& {Rose}, B. 2017,
  {Kbarbary/Sep: V1.0.2}, v1.0.2, Zenodo,  Zenodo,
  \dodoi{10.5281/zenodo.896928}

\bibitem[{{Barucci} {et~al.}(2018){Barucci}, {Perna}, {Popescu}, {Fornasier},
  {Doressoundiram}, {Lantz}, {Merlin}, {Fulchignoni}, {Dotto}, \&
  {Kanuchova}}]{Barucci2018}
{Barucci}, M.~A., {Perna}, D., {Popescu}, M., {et~al.} 2018, \mnras, 476, 4481,
  \dodoi{10.1093/mnras/sty532}

\bibitem[{{Belskaya} \& {Shevchenko}(2000)}]{Belskaya2000}
{Belskaya}, I.~N., \& {Shevchenko}, V.~G. 2000, Icarus, 147, 94,
  \dodoi{10.1006/icar.2000.6410}

\bibitem[{{Belskaya} {et~al.}(2003){Belskaya}, {Shevchenko}, {Kiselev},
  {Krugly}, {Shakhovskoy}, {Efimov}, {Gaftonyuk}, {Cellino}, \&
  {Gil-Hutton}}]{Belskaya2003}
{Belskaya}, I.~N., {Shevchenko}, V.~G., {Kiselev}, N.~N., {et~al.} 2003,
  \icarus, 166, 276, \dodoi{10.1016/j.icarus.2003.09.005}

\bibitem[{{Beniyama} {et~al.}(2022){Beniyama}, {Sako}, {Ohsawa}, {Takita},
  {Kobayashi}, {Okumura}, {Urakawa}, {Yoshikawa}, {Usui}, {Yoshida}, {Doi},
  {Niino}, {Shigeyama}, {Tanaka}, {Tominaga}, {Aoki}, {Arima}, {Arimatsu},
  {Kasuga}, {Kondo}, {Mori}, {Takahashi}, \& {Watanabe}}]{Beniyama2022}
{Beniyama}, J., {Sako}, S., {Ohsawa}, R., {et~al.} 2022, \pasj, 74, 877,
  \dodoi{10.1093/pasj/psac043}

\bibitem[{{Beniyama} {et~al.}(2023{\natexlab{a}}){Beniyama}, {Sekiguchi},
  {Kuroda}, {Arai}, {Ishibashi}, {Ishiguro}, {Yoshida}, {Senshu}, {Ootsubo},
  {Sako}, {Ohsawa}, {Takita}, {Geem}, \& {Bach}}]{Beniyama2023a}
{Beniyama}, J., {Sekiguchi}, T., {Kuroda}, D., {et~al.} 2023{\natexlab{a}},
  \pasj, 75, 297, \dodoi{10.1093/pasj/psac109}

\bibitem[{{Beniyama} {et~al.}(2023{\natexlab{b}}){Beniyama}, {Sako}, {Ohtsuka},
  {Sekiguchi}, {Ishiguro}, {Kuroda}, {Urakawa}, {Yoshida}, {Takumi}, {Maeda},
  {Takahashi}, {Takagi}, {Saito}, {Nakaoka}, {Saito}, {Ohshima}, {Imazawa},
  {Kagitani}, \& {Takita}}]{Beniyama2023b}
{Beniyama}, J., {Sako}, S., {Ohtsuka}, K., {et~al.} 2023{\natexlab{b}}, arXiv
  e-prints, arXiv:2306.15506, \dodoi{10.48550/arXiv.2306.15506}

\bibitem[{{Bertin} \& {Arnouts}(1996)}]{Bertin1996}
{Bertin}, E., \& {Arnouts}, S. 1996, \aaps, 117, 393,
  \dodoi{10.1051/aas:1996164}

\bibitem[{{Borisov} {et~al.}(2018){Borisov}, {Christou}, {Colas}, {Bagnulo},
  {Cellino}, \& {Dell'Oro}}]{Borisov2018b}
{Borisov}, G., {Christou}, A.~A., {Colas}, F., {et~al.} 2018, \aap, 618, A178,
  \dodoi{10.1051/0004-6361/201732466}

\bibitem[{{Bourdelle de Micas} {et~al.}(2022){Bourdelle de Micas}, {Fornasier},
  {Avdellidou}, {Delbo}, {van Belle}, {Ochner}, {Grundy}, \&
  {Moskovitz}}]{BourdelledeMicas2022}
{Bourdelle de Micas}, J., {Fornasier}, S., {Avdellidou}, C., {et~al.} 2022,
  \aap, 665, A83, \dodoi{10.1051/0004-6361/202244099}

\bibitem[{{Bowell} {et~al.}(1989){Bowell}, {Hapke}, {Domingue}, {Lumme},
  {Peltoniemi}, \& {Harris}}]{Bowell1989}
{Bowell}, E., {Hapke}, B., {Domingue}, D., {et~al.} 1989, in Asteroids II, ed.
  R.~P. {Binzel}, T.~{Gehrels}, \& M.~S. {Matthews} (Tucson, AZ: Univ. Arizona
  Press), 524--556

\bibitem[{{Cambioni} {et~al.}(2021){Cambioni}, {Delbo}, {Poggiali},
  {Avdellidou}, {Ryan}, {Deshapriya}, {Asphaug}, {Ballouz}, {Barucci},
  {Bennett}, {Bottke}, {Brucato}, {Burke}, {Cloutis}, {DellaGiustina}, {Emery},
  {Rozitis}, {Walsh}, \& {Lauretta}}]{Cambioni2021}
{Cambioni}, S., {Delbo}, M., {Poggiali}, G., {et~al.} 2021, \nat, 598, 49,
  \dodoi{10.1038/s41586-021-03816-5}

\bibitem[{{Chambers} {et~al.}(2016){Chambers}, {Magnier}, {Metcalfe},
  {Flewelling}, {Huber}, {Waters}, {Denneau}, {Draper}, {Farrow}, {Finkbeiner},
  {Holmberg}, {Koppenhoefer}, {Price}, {Rest}, {Saglia}, {Schlafly}, {Smartt},
  {Sweeney}, {Wainscoat}, {Burgett}, {Chastel}, {Grav}, {Heasley}, {Hodapp},
  {Jedicke}, {Kaiser}, {Kudritzki}, {Luppino}, {Lupton}, {Monet}, {Morgan},
  {Onaka}, {Shiao}, {Stubbs}, {Tonry}, {White}, {Ba{\~n}ados}, {Bell},
  {Bender}, {Bernard}, {Boegner}, {Boffi}, {Botticella}, {Calamida},
  {Casertano}, {Chen}, {Chen}, {Cole}, {Deacon}, {Frenk}, {Fitzsimmons},
  {Gezari}, {Gibbs}, {Goessl}, {Goggia}, {Gourgue}, {Goldman}, {Grant},
  {Grebel}, {Hambly}, {Hasinger}, {Heavens}, {Heckman}, {Henderson}, {Henning},
  {Holman}, {Hopp}, {Ip}, {Isani}, {Jackson}, {Keyes}, {Koekemoer}, {Kotak},
  {Le}, {Liska}, {Long}, {Lucey}, {Liu}, {Martin}, {Masci}, {McLean}, {Mindel},
  {Misra}, {Morganson}, {Murphy}, {Obaika}, {Narayan}, {Nieto-Santisteban},
  {Norberg}, {Peacock}, {Pier}, {Postman}, {Primak}, {Rae}, {Rai}, {Riess},
  {Riffeser}, {Rix}, {R{\"o}ser}, {Russel}, {Rutz}, {Schilbach}, {Schultz},
  {Scolnic}, {Strolger}, {Szalay}, {Seitz}, {Small}, {Smith}, {Soderblom},
  {Taylor}, {Thomson}, {Taylor}, {Thakar}, {Thiel}, {Thilker}, {Unger},
  {Urata}, {Valenti}, {Wagner}, {Walder}, {Walter}, {Watters}, {Werner},
  {Wood-Vasey}, \& {Wyse}}]{Chambers2016}
{Chambers}, K.~C., {Magnier}, E.~A., {Metcalfe}, N., {et~al.} 2016,
  arXiv:1612.05560, arXiv:1612.05560.
\newblock \doarXiv{1612.05560}

\bibitem[{{Colazo} {et~al.}(2023){Colazo}, {Fornari}, {Ciancia}, {Scotta},
  {Morales}, {Melia}, {Wilberger}, {Su{\'a}rez}, {Monteleone}, {Garc{\'\i}a},
  {Anzola}, {Santos}, {Mottino}, \& {Colazo}}]{Colazo2023}
{Colazo}, M., {Fornari}, C., {Ciancia}, G., {et~al.} 2023, Minor Planet
  Bulletin, 50, 235

\bibitem[{{Delbo'} {et~al.}(2017){Delbo'}, {Walsh}, {Bolin}, {Avdellidou}, \&
  {Morbidelli}}]{Delbo2017}
{Delbo'}, M., {Walsh}, K., {Bolin}, B., {Avdellidou}, C., \& {Morbidelli}, A.
  2017, Science, 357, 1026, \dodoi{10.1126/science.aam6036}

\bibitem[{{DeMeo} {et~al.}(2019){DeMeo}, {Polishook}, {Carry}, {Burt}, {Hsieh},
  {Binzel}, {Moskovitz}, \& {Burbine}}]{DeMeo2019}
{DeMeo}, F.~E., {Polishook}, D., {Carry}, B., {et~al.} 2019, \icarus, 322, 13,
  \dodoi{10.1016/j.icarus.2018.12.016}

\bibitem[{{Fenucci} {et~al.}(2023){Fenucci}, {Novakovi{\'c}}, \&
  {Mar{\v{c}}eta}}]{Fenucci2023b}
{Fenucci}, M., {Novakovi{\'c}}, B., \& {Mar{\v{c}}eta}, D. 2023, \aap, 675,
  A134, \dodoi{10.1051/0004-6361/202346160}

\bibitem[{{Fenucci} {et~al.}(2021){Fenucci}, {Novakovi{\'c}},
  {Vokrouhlick{\'y}}, \& {Weryk}}]{Fenucci2021}
{Fenucci}, M., {Novakovi{\'c}}, B., {Vokrouhlick{\'y}}, D., \& {Weryk}, R.~J.
  2021, \aap, 647, A61, \dodoi{10.1051/0004-6361/202039628}

\bibitem[{{Fowler} \& {Chillemi}(1992)}]{Fowler1992}
{Fowler}, J.~W., \& {Chillemi}, J.~R. 1992, Phillips Lab. Tech. Rep., 2049, 17

\bibitem[{{Franco} {et~al.}(2023){Franco}, {Marchini}, {Iozzi}, {Galli},
  {Montigiani}, {Mannucci}, {Scarfi}, {Coffano}, {Marinello}, {Mattei},
  {Ruocco}, \& {Baj}}]{Franco2023b}
{Franco}, L., {Marchini}, A., {Iozzi}, M., {et~al.} 2023, Minor Planet
  Bulletin, 50, 173

\bibitem[{{Ginsburg} {et~al.}(2019){Ginsburg}, {Sip{\H{o}}cz}, {Brasseur},
  {Cowperthwaite}, {Craig}, {Deil}, {Guillochon}, {Guzman}, {Liedtke}, {Lian
  Lim}, {Lockhart}, {Mommert}, {Morris}, {Norman}, {Parikh}, {Persson},
  {Robitaille}, {Segovia}, {Singer}, {Tollerud}, {de Val-Borro}, {Valtchanov},
  {Woillez}, {Astroquery Collaboration}, \& {a subset of astropy
  Collaboration}}]{Ginsburg2019}
{Ginsburg}, A., {Sip{\H{o}}cz}, B.~M., {Brasseur}, C.~E., {et~al.} 2019, \aj,
  157, 98, \dodoi{10.3847/1538-3881/aafc33}

\bibitem[{{Harris} \& {Lupishko}(1989)}]{Harris1989}
{Harris}, A.~W., \& {Lupishko}, D.~F. 1989, in Asteroids II, ed. R.~P.
  {Binzel}, T.~{Gehrels}, \& M.~S. {Matthews}, 39--53

\bibitem[{{Harris} {et~al.}(2020){Harris}, {Millman}, {van der Walt},
  {Gommers}, {Virtanen}, {Cournapeau}, {Wieser}, {Taylor}, {Berg}, {Smith},
  {Kern}, {Picus}, {Hoyer}, {van Kerkwijk}, {Brett}, {Haldane}, {del R{\'\i}o},
  {Wiebe}, {Peterson}, {G{\'e}rard-Marchant}, {Sheppard}, {Reddy}, {Weckesser},
  {Abbasi}, {Gohlke}, \& {Oliphant}}]{Harris2020}
{Harris}, C.~R., {Millman}, K.~J., {van der Walt}, S.~J., {et~al.} 2020, \nat,
  585, 357, \dodoi{10.1038/s41586-020-2649-2}

\bibitem[{{Hasegawa} {et~al.}(2021){Hasegawa}, {Marsset}, {DeMeo}, {Bus},
  {Geem}, {Ishiguro}, {Im}, {Kuroda}, \& {Vernazza}}]{Hasegawa2021b}
{Hasegawa}, S., {Marsset}, M., {DeMeo}, F.~E., {et~al.} 2021, \apjl, 916, L6,
  \dodoi{10.3847/2041-8213/ac0f05}

\bibitem[{Hunter(2007)}]{Hunter2007}
Hunter, J.~D. 2007, Computing in Science \& Engineering, 9, 90,
  \dodoi{10.1109/MCSE.2007.55}

\bibitem[{{Ieva} {et~al.}(2022){Ieva}, {Arcoverde}, {Rond{\'o}n}, {Giunta},
  {Dotto}, {Lazzaro}, {Mazzotta Epifani}, {Perna}, {Fanasca}, {Rodrigues},
  {Monteiro}, {Medeiros}, {Silva-Cabrera}, \& {Di Paola}}]{Ieva2022}
{Ieva}, S., {Arcoverde}, P., {Rond{\'o}n}, E., {et~al.} 2022, \mnras, 513,
  3104, \dodoi{10.1093/mnras/stac1117}

\bibitem[{{Jehin} {et~al.}(2011){Jehin}, {Gillon}, {Queloz}, {Magain},
  {Manfroid}, {Chantry}, {Lendl}, {Hutsem{\'e}kers}, \& {Udry}}]{Jehin2011}
{Jehin}, E., {Gillon}, M., {Queloz}, D., {et~al.} 2011, The Messenger, 145, 2

\bibitem[{{Juri{\'c}} {et~al.}(2002){Juri{\'c}}, {Ivezi{\'c}}, {Lupton},
  {Quinn}, {Tabachnik}, {Fan}, {Gunn}, {Hennessy}, {Knapp}, {Munn}, {Pier},
  {Rockosi}, {Schneider}, {Brinkmann}, {Csabai}, \& {Fukugita}}]{Juric2002}
{Juri{\'c}}, M., {Ivezi{\'c}}, {\v{Z}}., {Lupton}, R.~H., {et~al.} 2002, \aj,
  124, 1776, \dodoi{10.1086/341950}

\bibitem[{{Kole^^c5^^84czuk} {et~al.}(2023){Kole^^c5^^84czuk}, {Kwiatkowski},
  {Kami^^c5^^84ska}, \& {Kaminski}}]{Kolenczuk2023}
{Kole^^c5^^84czuk}, P., {Kwiatkowski}, T., {Kami^^c5^^84ska}, M., \&
  {Kaminski}, K. 2023, in Asteroids, Comets, and Meteors: ACM 2023

\bibitem[{{Kurita} {et~al.}(2020){Kurita}, {Kino}, {Iwamuro}, {Ohta}, {Nogami},
  {Izumiura}, {Yoshida}, {Matsubayashi}, {Kuroda}, {Nakatani}, {Yamamoto},
  {Tsutsui}, {Iribe}, {Jikuya}, {Ohtani}, {Shibata}, {Takahashi}, {Tokoro},
  {Maihara}, \& {Nagata}}]{Kurita2020}
{Kurita}, M., {Kino}, M., {Iwamuro}, F., {et~al.} 2020, \pasj, 72, 48,
  \dodoi{10.1093/pasj/psaa036}

\bibitem[{{Lang} {et~al.}(2010){Lang}, {Hogg}, {Mierle}, {Blanton}, \&
  {Roweis}}]{Lang2010}
{Lang}, D., {Hogg}, D.~W., {Mierle}, K., {Blanton}, M., \& {Roweis}, S. 2010,
  \aj, 139, 1782, \dodoi{10.1088/0004-6256/139/5/1782}

\bibitem[{{Lee} {et~al.}(2017){Lee}, {Moon}, {Kim}, {Kim}, {Durech}, {Choi},
  {Oh}, {Park}, {Roh}, {Yim}, {Cha}, \& {Lee}}]{Lee2017}
{Lee}, H.-J., {Moon}, H.-K., {Kim}, M.-J., {et~al.} 2017, Journal of Korean
  Astronomical Society, 50, 41, \dodoi{10.5303/JKAS.2017.50.3.41}

\bibitem[{{Lee} {et~al.}(2022){Lee}, {Kim}, {Marciniak}, {Kim}, {Moon}, {Choi},
  {Zo{\l}a}, {Chatelain}, {Lister}, {Gomez}, {Greenstreet}, {P{\'a}l},
  {Szak{\'a}ts}, {Erasmus}, {Lees}, {Janse van Rensburg}, {Og{\l}oza},
  {Dr{\'o}{\.z}d{\.z}}, {{\.Z}ejmo}, {Kami{\'n}ski}, {Kami{\'n}ska}, {Duffard},
  {Roh}, {Yim}, {Kim}, {Mottola}, {Yoshida}, {Reichart}, {Sonbas}, {Caton},
  {Kaplan}, {Erece}, \& {Yang}}]{Lee2022}
{Lee}, H.~J., {Kim}, M.~J., {Marciniak}, A., {et~al.} 2022, \aap, 661, L3,
  \dodoi{10.1051/0004-6361/202243442}

\bibitem[{{Licandro} {et~al.}(2023){Licandro}, {Popescu}, {Tatsumi}, {Alarcon},
  {Serra-Ricart}, {Medeiros}, {Morate}, {Tinaut-Ruano}, \& {de
  Le{\'o}n}}]{Licandro2023}
{Licandro}, J., {Popescu}, M., {Tatsumi}, E., {et~al.} 2023, \mnras, 521, 3784,
  \dodoi{10.1093/mnras/stad708}

\bibitem[{{Lomb}(1976)}]{Lomb1976}
{Lomb}, N.~R. 1976, \apss, 39, 447, \dodoi{10.1007/BF00648343}

\bibitem[{{Mahlke} {et~al.}(2021){Mahlke}, {Carry}, \& {Denneau}}]{Mahlke2021}
{Mahlke}, M., {Carry}, B., \& {Denneau}, L. 2021, \icarus, 354, 114094,
  \dodoi{10.1016/j.icarus.2020.114094}

\bibitem[{{Mahlke} {et~al.}(2022){Mahlke}, {Carry}, \& {Mattei}}]{Mahlke2022}
{Mahlke}, M., {Carry}, B., \& {Mattei}, P.~A. 2022, \aap, 665, A26,
  \dodoi{10.1051/0004-6361/202243587}

\bibitem[{{Martikainen} {et~al.}(2021){Martikainen}, {Muinonen},
  {Penttil{\"a}}, {Cellino}, \& {Wang}}]{Martikainen2021}
{Martikainen}, J., {Muinonen}, K., {Penttil{\"a}}, A., {Cellino}, A., \&
  {Wang}, X.~B. 2021, \aap, 649, A98, \dodoi{10.1051/0004-6361/202039796}

\bibitem[{{McCully} {et~al.}(2018){McCully}, {Crawford}, {Kovacs}, {Tollerud},
  {Betts}, {Bradley}, {Craig}, {Turner}, {Streicher}, {Sipocz}, {Robitaille},
  \& {Deil}}]{McCully2018}
{McCully}, C., {Crawford}, S., {Kovacs}, G., {et~al.} 2018,
  {Astropy/Astroscrappy: V1.0.5 Zenodo Release}, v1.0.5, Zenodo,  Zenodo,
  \dodoi{10.5281/zenodo.1482019}

\bibitem[{{Mommert}(2017)}]{Mommert2017}
{Mommert}, M. 2017, Astronomy and Computing, 18, 47,
  \dodoi{10.1016/j.ascom.2016.11.002}

\bibitem[{{Mommert} {et~al.}(2014{\natexlab{a}}){Mommert}, {Hora},
  {Farnocchia}, {Chesley}, {Vokrouhlick{\'y}}, {Trilling}, {Mueller}, {Harris},
  {Smith}, \& {Fazio}}]{Mommert2014a}
{Mommert}, M., {Hora}, J.~L., {Farnocchia}, D., {et~al.} 2014{\natexlab{a}},
  \apj, 786, 148, \dodoi{10.1088/0004-637X/786/2/148}

\bibitem[{{Mommert} {et~al.}(2014{\natexlab{b}}){Mommert}, {Farnocchia},
  {Hora}, {Chesley}, {Trilling}, {Chodas}, {Mueller}, {Harris}, {Smith}, \&
  {Fazio}}]{Mommert2014b}
{Mommert}, M., {Farnocchia}, D., {Hora}, J.~L., {et~al.} 2014{\natexlab{b}},
  \apjl, 789, L22, \dodoi{10.1088/2041-8205/789/1/L22}

\bibitem[{{Muinonen} {et~al.}(2010){Muinonen}, {Belskaya}, {Cellino},
  {Delb{\`o}}, {Levasseur-Regourd}, {Penttil{\"a}}, \&
  {Tedesco}}]{Muinonen2010}
{Muinonen}, K., {Belskaya}, I.~N., {Cellino}, A., {et~al.} 2010, \icarus, 209,
  542, \dodoi{10.1016/j.icarus.2010.04.003}

\bibitem[{Oliphant(2015)}]{Oliphant2015}
Oliphant, T.~E. 2015, Guide to NumPy, 2nd edn. (North Charleston, SC, USA:
  CreateSpace Independent Publishing Platform)

\bibitem[{{Petrov} {et~al.}(2018){Petrov}, {Vasil'ev}, {Kuteeva}, \&
  {Sokolov}}]{Petrov2018}
{Petrov}, N.~A., {Vasil'ev}, A.~A., {Kuteeva}, G.~A., \& {Sokolov}, L.~L. 2018,
  Solar System Research, 52, 326, \dodoi{10.1134/S0038094618040032}

\bibitem[{{Polishook} {et~al.}(2017){Polishook}, {Jacobson}, {Morbidelli}, \&
  {Aharonson}}]{Polishook2017a}
{Polishook}, D., {Jacobson}, S.~A., {Morbidelli}, A., \& {Aharonson}, O. 2017,
  Nature Astronomy, 1, 0179, \dodoi{10.1038/s41550-017-0179}

\bibitem[{{Popescu} {et~al.}(2012){Popescu}, {Birlan}, \&
  {Nedelcu}}]{Popescu2012}
{Popescu}, M., {Birlan}, M., \& {Nedelcu}, D.~A. 2012, \aap, 544, A130,
  \dodoi{10.1051/0004-6361/201219584}

\bibitem[{{Pravec} \& {Harris}(2007)}]{Pravec2007}
{Pravec}, P., \& {Harris}, A.~W. 2007, Icarus, 190, 250,
  \dodoi{10.1016/j.icarus.2007.02.023}

\bibitem[{{Pravec} {et~al.}(2012){Pravec}, {Harris}, {Ku{\v{s}}nir{\'a}k},
  {Gal{\'a}d}, \& {Hornoch}}]{Pravec2012}
{Pravec}, P., {Harris}, A.~W., {Ku{\v{s}}nir{\'a}k}, P., {Gal{\'a}d}, A., \&
  {Hornoch}, K. 2012, \icarus, 221, 365, \dodoi{10.1016/j.icarus.2012.07.026}

\bibitem[{{Pravec} {et~al.}(2005){Pravec}, {Harris}, {Scheirich},
  {Ku{\v{s}}nir{\'a}k}, {{\v{S}}arounov{\'a}}, {Hergenrother}, {Mottola},
  {Hicks}, {Masi}, {Krugly}, {Shevchenko}, {Nolan}, {Howell}, {Kaasalainen},
  {Gal{\'a}d}, {Brown}, {DeGraff}, {Lambert}, {Cooney}, \&
  {Foglia}}]{Pravec2005}
{Pravec}, P., {Harris}, A.~W., {Scheirich}, P., {et~al.} 2005, Icarus, 173,
  108, \dodoi{10.1016/j.icarus.2004.07.021}

\bibitem[{{Pravec} {et~al.}(2014){Pravec}, {Scheirich}, {{\v{D}}urech},
  {Pollock}, {Ku{\v{s}}nir{\'a}k}, {Hornoch}, {Gal{\'a}d}, {Vokrouhlick{\'y}},
  {Harris}, {Jehin}, {Manfroid}, {Opitom}, {Gillon}, {Colas}, {Oey},
  {Vra{\v{s}}til}, {Reichart}, {Ivarsen}, {Haislip}, \&
  {LaCluyze}}]{Pravec2014}
{Pravec}, P., {Scheirich}, P., {{\v{D}}urech}, J., {et~al.} 2014, \icarus, 233,
  48, \dodoi{10.1016/j.icarus.2014.01.026}

\bibitem[{{Rayner} {et~al.}(2003){Rayner}, {Toomey}, {Onaka}, {Denault},
  {Stahlberger}, {Vacca}, {Cushing}, \& {Wang}}]{Rayner2003}
{Rayner}, J.~T., {Toomey}, D.~W., {Onaka}, P.~M., {et~al.} 2003, \pasp, 115,
  362, \dodoi{10.1086/367745}

\bibitem[{{Reback} {et~al.}(2021){Reback}, {jbrockmendel}, {McKinney}, {Van den
  Bossche}, {Augspurger}, {Cloud}, {Hawkins}, {gfyoung}, {Roeschke}, {Sinhrks},
  {Klein}, {Petersen}, {Tratner}, {She}, {Ayd}, {Hoefler}, {Naveh}, {Garcia},
  {Schendel}, {Hayden}, {Saxton}, {Darbyshire}, {Shadrach}, {Gorelli}, {Li},
  {Zeitlin}, {Jancauskas}, {McMaster}, {Battiston}, \& {Seabold}}]{Reback2021}
{Reback}, J., {jbrockmendel}, {McKinney}, W., {et~al.} 2021,
  {pandas-dev/pandas: Pandas 1.3.4}, v1.3.4, Zenodo,  Zenodo,
  \dodoi{10.5281/zenodo.5574486}

\bibitem[{{Reddy} {et~al.}(2015){Reddy}, {Gary}, {Sanchez}, {Takir}, {Thomas},
  {Hardersen}, {Ogmen}, {Benni}, {Kaye}, {Gregorio}, {Garlitz}, {Polishook},
  {Le Corre}, \& {Nathues}}]{Reddy2015}
{Reddy}, V., {Gary}, B.~L., {Sanchez}, J.~A., {et~al.} 2015, \apj, 811, 65,
  \dodoi{10.1088/0004-637X/811/1/65}

\bibitem[{{Reddy} {et~al.}(2016){Reddy}, {Sanchez}, {Bottke}, {Thirouin},
  {Rivera-Valentin}, {Kelley}, {Ryan}, {Cloutis}, {Tegler}, {Ryan}, {Taylor},
  {Richardson}, {Moskovitz}, \& {Le Corre}}]{Reddy2016}
{Reddy}, V., {Sanchez}, J.~A., {Bottke}, W.~F., {et~al.} 2016, \aj, 152, 162,
  \dodoi{10.3847/0004-6256/152/6/162}

\bibitem[{{Rond{\'o}n} {et~al.}(2019){Rond{\'o}n}, {Arcoverde}, {Monteiro},
  {Medeiros}, {Navas}, {Lazzaro}, {Carvano}, \& {Rodrigues}}]{Rondon2019}
{Rond{\'o}n}, E., {Arcoverde}, P., {Monteiro}, F., {et~al.} 2019, \mnras, 484,
  2499, \dodoi{10.1093/mnras/stz024}

\bibitem[{{Rond{\'o}n} {et~al.}(2022){Rond{\'o}n}, {Lazzaro}, {Carvano},
  {Monteiro}, {Arcoverde}, {Evangelista}, {Michimani}, {Mesquita}, \&
  {Rodrigues}}]{Rondon2022}
{Rond{\'o}n}, E., {Lazzaro}, D., {Carvano}, J., {et~al.} 2022, \icarus, 372,
  114723, \dodoi{10.1016/j.icarus.2021.114723}

\bibitem[{Rubincam(2000)}]{Rubincam2000}
Rubincam, D.~P. 2000, Icarus, 148, 2, \dodoi{10.1006/icar.2000.6485}

\bibitem[{{Scargle}(1982)}]{Scargle1982}
{Scargle}, J.~D. 1982, \apj, 263, 835, \dodoi{10.1086/160554}

\bibitem[{{Shevchenko} {et~al.}(2016){Shevchenko}, {Belskaya}, {Muinonen},
  {Penttil{\"a}}, {Krugly}, {Velichko}, {Chiorny}, {Slyusarev}, {Gaftonyuk}, \&
  {Tereschenko}}]{Shevchenko2016}
{Shevchenko}, V.~G., {Belskaya}, I.~N., {Muinonen}, K., {et~al.} 2016, \planss,
  123, 101, \dodoi{10.1016/j.pss.2015.11.007}

\bibitem[{{Terai} {et~al.}(2013){Terai}, {Urakawa}, {Takahashi}, {Yoshida},
  {Oshima}, {Aratani}, {Hoshi}, {Sato}, {Ushioda}, \& {Oasa}}]{Terai2013}
{Terai}, T., {Urakawa}, S., {Takahashi}, J., {et~al.} 2013, \aap, 559, A106,
  \dodoi{10.1051/0004-6361/201322158}

\bibitem[{{Thirouin} {et~al.}(2016){Thirouin}, {Moskovitz}, {Binzel},
  {Christensen}, {DeMeo}, {Person}, {Polishook}, {Thomas}, {Trilling},
  {Willman}, {Hinkle}, {Burt}, {Avner}, \& {Aceituno}}]{Thirouin2016}
{Thirouin}, A., {Moskovitz}, N., {Binzel}, R.~P., {et~al.} 2016, \aj, 152, 163,
  \dodoi{10.3847/0004-6256/152/6/163}

\bibitem[{{Thirouin} {et~al.}(2018){Thirouin}, {Moskovitz}, {Binzel},
  {Christensen}, {DeMeo}, {Person}, {Polishook}, {Thomas}, {Trilling},
  {Willman}, {Burt}, {Hinkle}, \& {Pugh}}]{Thirouin2018}
{Thirouin}, A., {Moskovitz}, N.~A., {Binzel}, R.~P., {et~al.} 2018, \apjs, 239,
  4, \dodoi{10.3847/1538-4365/aae1b0}

\bibitem[{{Tonry} {et~al.}(2012){Tonry}, {Stubbs}, {Lykke}, {Doherty},
  {Shivvers}, {Burgett}, {Chambers}, {Hodapp}, {Kaiser}, {Kudritzki},
  {Magnier}, {Morgan}, {Price}, \& {Wainscoat}}]{Tonry2012}
{Tonry}, J.~L., {Stubbs}, C.~W., {Lykke}, K.~R., {et~al.} 2012, \apj, 750, 99,
  \dodoi{10.1088/0004-637X/750/2/99}

\bibitem[{{Usui} {et~al.}(2013){Usui}, {Kasuga}, {Hasegawa}, {Ishiguro},
  {Kuroda}, {M{\"u}ller}, {Ootsubo}, \& {Matsuhara}}]{Usui2013}
{Usui}, F., {Kasuga}, T., {Hasegawa}, S., {et~al.} 2013, \apj, 762, 56,
  \dodoi{10.1088/0004-637X/762/1/56}

\bibitem[{{van Dokkum}(2001)}]{vanDokkum2001}
{van Dokkum}, P.~G. 2001, \pasp, 113, 1420, \dodoi{10.1086/323894}

\bibitem[{{VanderPlas}(2018)}]{VanderPlas2018}
{VanderPlas}, J.~T. 2018, \apjs, 236, 16, \dodoi{10.3847/1538-4365/aab766}

\bibitem[{{Virtanen} {et~al.}(2020){Virtanen}, {Gommers}, {Oliphant},
  {Haberland}, {Reddy}, {Cournapeau}, {Burovski}, {Peterson}, {Weckesser},
  {Bright}, {van der Walt}, {Brett}, {Wilson}, {Millman}, {Mayorov}, {Nelson},
  {Jones}, {Kern}, {Larson}, {Carey}, {Polat}, {Feng}, {Moore}, {VanderPlas},
  {Laxalde}, {Perktold}, {Cimrman}, {Henriksen}, {Quintero}, {Harris},
  {Archibald}, {Ribeiro}, {Pedregosa}, {van Mulbregt}, \& {SciPy 1. 0
  Contributors}}]{Virtanen2020}
{Virtanen}, P., {Gommers}, R., {Oliphant}, T.~E., {et~al.} 2020, Nature
  Methods, 17, 261, \dodoi{10.1038/s41592-019-0686-2}

\bibitem[{{Vokrouhlick{\'y}}(1998)}]{Vokrouhlicky1998}
{Vokrouhlick{\'y}}, D. 1998, \aap, 335, 1093

\bibitem[{{Willmer}(2018)}]{Willmer2018}
{Willmer}, C. N.~A. 2018, \apjs, 236, 47, \dodoi{10.3847/1538-4365/aabfdf}

\end{thebibliography}
\end{document}